\documentclass[prb,reprint,showkeys]{revtex4-2}
\usepackage{setspace, parskip, xcolor, breakcites, hyphenat, graphicx, epstopdf, amsmath, amssymb, amsfonts, array,
	verbatim, url, subcaption, booktabs, multirow, etoolbox,siunitx, etoolbox, tikz,stackengine, textcomp,
	gensymb, siunitx,tabularx, array, xfp,overpic}
\def\stripzero#1{\expandafter\stripzerohelp#1}

\def\stripzerohelp#1{\ifx 0#1\expandafter\stripzerohelp\else#1\fi}
\newcolumntype{H}{>{\setbox0=\hbox\bgroup}c<{\egroup}@{}}
\usepackage[toc,page,titletoc]{appendix}
\usepackage[inline]{enumitem}
\usepackage[version=4]{mhchem}
\usepackage[T1]{fontenc}
\usetikzlibrary{shadings}
\definecolor{srm}{HTML}{034da1}
\definecolor{cof}{RGB}{219,144,71}
\definecolor{pur}{RGB}{186,146,162}
\definecolor{greeo}{RGB}{91,173,69}
\definecolor{greet}{RGB}{52,111,72}
\definecolor{mplc0}{HTML}{1f77b4}
\definecolor{mplc1}{HTML}{ff7f0e}
\definecolor{mplc2}{HTML}{2e7d32}
\definecolor{mplc3}{HTML}{d32f2f}
\definecolor{mplc4}{HTML}{9467bd}
\definecolor{mplc5}{HTML}{8c564b}
\definecolor{mplc6}{HTML}{e377c2}
\definecolor{mplc7}{HTML}{7f7f7f}
\definecolor{mplc8}{HTML}{bcbd22}
\definecolor{mplc9}{HTML}{17becf}
\definecolor{Astral}{HTML}{1F77B4}
\definecolor{BG80}{HTML}{37474f}
\definecolor{Green}{HTML}{4CA540}
\definecolor{Red}{HTML}{F44336}
\usepackage[colorlinks=true, linkcolor=mplc3, urlcolor=mplc3, filecolor=mplc0, citecolor=mplc2,
	pdfstartview=FitV, pdftitle={}, pdfauthor={}, pdfsubject={}, pdfkeywords={}, pdfpagemode={},
	bookmarksopen=true, breaklinks]{hyperref}
\usepackage[capitalise,nameinlink]{cleveref}
\crefname{supp}{Supplement}{Supplements}
\PassOptionsToPackage{hyphens}{url}
\newcommand{\etal}{\textit{et al.~}}

\newcommand{\figref}[1]{Fig. \ref{#1}}

\newcommand{\tabref}[1]{Table~\ref{#1}}
\newcommand{\tis}{\ce{TiS_{2-x}}}
\newcommand{\ptis}{\ce{TiS2}}
\graphicspath{{./figures}, {./figsi}, {./figprl}}
\date{\today}

\makeatletter
\def\@email#1#2{%
	\endgroup
	\patchcmd{\titleblock@produce}
	{\frontmatter@RRAPformat}
	{\frontmatter@RRAPformat{\produce@RRAP{*#1\href{mailto:#2}{#2}}}\frontmatter@RRAPformat}
	{}{}
}%
\makeatother
\begin{document}
\title{Geometric Percolation Threshold Defines Half-Metallic Window in Vacancy-Doped \ptis}
\author{Shrestha Dutta}
\author{Rudra Banerjee*}\email{rudrab@srmist.edu.in}
\affiliation{Department of Physics and Nanotechnology, SRM Institute of Science and Technology, Kattankulathur, Tamil
	Nadu, 603203, India}

\begin{abstract}
	Defect engineering of two-dimensional materials routinely produces local magnetic moments, yet itinerant half-metallic
	ferromagnetism remains elusive---experiments frequently yield paramagnetic insulators. We resolve this paradox for vacancy-doped
	monolayer $1T$-\ptis~by demonstrating that the insulator-to-half-metal transition is governed by universal geometric percolation
	of the defect network, extending the percolation framework established for three-dimensional diluted magnetic semiconductors into
	the 2D vacancy-doped regime. Half-metallicity emerges via a two-step mechanism: crystal-field symmetry breaking
	($O_h \to C_{4v}$) selectively stabilizes the Ti $3d_{z^2}$ orbital, generating robust local moments ($0.94~\mu_B$), but
	spin-polarized transport requires these moments to form a spanning cluster. At critical vacancy concentration $x_c \approx
		12.5\%$, a percolation transition drives the majority-spin impurity band from flat, localized levels ($W < 0.1$~eV) to a
	dispersive 1.5~eV-wide band with 100\% spin polarization and a minority-spin gap of 1.0~eV. Finite-size scaling yields a Fisher
	exponent $\tau = 2.09 \pm 0.03$, confirmed by fractal scaling of \textit{ab initio} charge densities
	($\tau_{\text{eff}}^{\text{DFT}} = 1.87 \pm 0.26$), placing the transition in the 2D percolation universality class. The
	percolation mechanism is independently corroborated by a striking supercell-size effect: at identical concentration, $2\times2$
	cells yield antiferromagnetic order while $4\times4$ cells mandate ferromagnetism, reflecting the presence or absence of a
	spanning cluster. We estimate a Curie temperature exceeding 300~K from the exchange coupling, and identify a geometric jamming
	instability at $x > 20\%$ that fragments the network. These results define a narrow functional window ($11\% < x < 15\%$) for
	half-metallic operation and establish geometric connectivity as a quantitative design principle for defect-engineered 2D
	spintronics.
\end{abstract}
\keywords{Percolation, Half-metals, Transition metal dichalcogenides, Spintronics, Density functional theory, 2D materials}
\maketitle

\section{Introduction}
\label{sec:introduction}

Half-metallic ferromagnets---materials that are metallic for one spin channel and insulating for the other---have been a central
target of spintronics since their theoretical prediction by de Groot \etal in 1983~\cite{deGroot1983}. The promise of 100\% spin
polarization at the Fermi level would enable effective spin injection, a prerequisite for spin-transfer torque memories,
spin filters, and energy-efficient logic~\cite{Wolf_2001, Ahn_2020}. Despite decades of effort in bulk Heusler alloys and
complex oxides, realizing robust half-metallicity in technologically viable geometries remains an open
challenge~\cite{Liu_2020}. The recent discovery of intrinsic ferromagnetism in atomically thin van der Waals crystals---\ce{CrI3}
and \ce{Cr2Ge2Te6}~\cite{Huang2017, Gong2017}---has opened a new frontier by demonstrating that magnetic order can survive in
the two-dimensional (2D) limit, reviving interest in 2D platforms for spintronic applications~\cite{Gibertini2019, Manzeli_2017}.

A particularly appealing strategy is defect engineering of the far more abundant non-magnetic 2D materials. Transition metal
dichalcogenides (TMDCs) offer atomic-scale thickness, direct bandgaps, strong spin-orbit coupling, and spin-valley
locking~\cite{Manzeli_2017, Liu_2020}, yet they lack intrinsic magnetism. Introducing chalcogen vacancies has emerged as a
leading route to induce local magnetic moments and, potentially, half-metallic order~\cite{Cai2015, Hossen2024, Noor2024}.
However, a persistent gap separates theory from experiment: density functional theory (DFT) calculations routinely predict
stable magnetic ground states in vacancy-doped TMDCs, whereas experiments frequently observe paramagnetic insulators or
negligible spin signals~\cite{Gaikwad2025, Hotger2023}. Recent spin-resolved measurements on single sulfur vacancies in
\ce{MoS2} confirm that individual defects carry spin, yet long-range magnetic order fails to
emerge~\cite{Hotger2023}. This discrepancy highlights a fundamental conceptual gap: the existence of local moments is a
necessary but insufficient condition for itinerant magnetism.

The missing ingredient is geometric connectivity. This insight has a well-established precedent in three-dimensional diluted
magnetic semiconductors (DMSs). In the prototypical system (Ga,Mn)As, Dietl \etal~\cite{Dietl2000} showed that carrier-mediated
ferromagnetism requires a minimum concentration of magnetic dopants, and Bergqvist \etal~\cite{Bergqvist2004} demonstrated
explicitly that this threshold corresponds to the percolation of magnetic clusters---a geometric phase transition that gates the
onset of long-range order. Subsequent work by Sato \etal~\cite{Sato2010} established that the magnetic phase diagram of an
entire class of DMSs is governed by percolation geometry. The implications are profound: ferromagnetic order is not solely
determined by the strength of exchange coupling, but by whether the magnetic sites form a connected network capable of
sustaining collective transport. Stiller and Esquinazi~\cite{Stiller2023} extended this reasoning to defect-induced
ferromagnetism in \ce{TiO2}, showing that the percolation threshold governs the onset of quasi-2D magnetic order. Similar
percolation-driven transitions have been identified in the metal-insulator transition of
\ce{LaMnO3}~\cite{Sherafati2016}. Yet despite its predictive success in 3D systems, this geometric framework has never been
rigorously applied to 2D vacancy-doped materials, where the reduced dimensionality fundamentally alters both the percolation
universality class and the physics of magnetic ordering~\cite{stauffer1994introduction, saberi2015recent}.

Monolayer $1T$-\ptis~provides the ideal testbed to isolate the geometric mechanism from the confounding effects of moment
quenching. In many Group-VI TMDCs---most notably $2H$-\ce{MoS2} and \ce{WS2}---strong inward lattice relaxation around a
sulfur vacancy reconstructs the local bonding environment, quenching the magnetic moment and rendering the defect magnetically
inert~\cite{Gaikwad2025, Hossen2024}. In \ptis, the outcome is fundamentally different. The removal of a sulfur atom from the
$1T$ octahedral framework reduces the local coordination from octahedral ($O_h$) to square-pyramidal ($C_{4v}$), lifting the
crystal-field degeneracy of the Ti $3d$ manifold~\cite{Zhang_2024}. The $d_{z^2}$ orbital, oriented directly toward the
vacancy, is selectively stabilized by the loss of antibonding overlap with the missing anion, creating a localized trap state
within the band gap. This orbital-selective symmetry breaking ensures that each sulfur vacancy consistently generates a
paramagnetic Ti$^{3+}$ ($d^1$) center with a robust moment of $\sim 1~\mu_B$~\cite{Wu_2022}. With the local moment protected by
crystal-field symmetry rather than being a fragile consequence of weak exchange splitting, the remaining question is purely
geometric: at what critical density do these isolated magnetic polarons percolate into a system-spanning conducting network?

Here, we answer this question quantitatively. Using DFT combined with continuum percolation analysis on \textit{ab initio}
charge densities and large-scale tight-binding simulations, we demonstrate that the insulator-to-half-metal transition in
\tis~is driven by universal geometric percolation. We report three principal findings. First, robust half-metallic
ferromagnetism with 100\% spin polarization emerges at $x_c \approx 12.5\%$ vacancy concentration, precisely synchronized with
the formation of a giant spanning cluster ($P_\infty \approx 30\%$). Second, finite-size scaling on $80 \times 80$ lattices
yields a Fisher exponent $\tau = 2.09 \pm 0.03$, in excellent agreement with the exact 2D percolation universality class
($\tau_{\text{theory}} = 187/91 \approx 2.05$), confirming that the electronic transition belongs to a well-defined universality
class. Third, we identify a geometric jamming instability at $x > 20\%$ that fragments the percolating network and collapses the
half-metallic phase, imposing a hard upper bound on the functional doping range. Together, these results define a narrow
operational window ($11\% < x < 15\%$) for defect-engineered half-metallicity in \ptis~and establish geometric connectivity as
a quantitative design principle for 2D spintronics.

\section{Computational Framework}
\label{sec:computational_method}

\subsection{Density Functional Theory}
\label{sub:dft}

Electronic structure calculations were performed using spin-polarized DFT as implemented in the Vienna Ab initio Simulation
Package (\textsc{vasp})~\cite{vasp1,vasp2}. Exchange-correlation effects were treated within the Generalized Gradient
Approximation (GGA) parameterized by Perdew, Burke, and Ernzerhof (PBE)~\cite{pbe}. To correct the self-interaction error
inherent in $3d$ transition metal chalcogenides, we applied a Hubbard-$U$ correction ($U_{\text{eff}} = 6.7$~eV) to the Ti
$d$-orbitals within the rotationally invariant Dudarev formalism~\cite{Dudarev_1998}. This parameter was calibrated by
reproducing the experimental monolayer band gap ($E_g = 1$~eV) reported from angle-resolved photoemission spectroscopy
(ARPES)~\cite{Yanagizawa_2025} and cross-validated against hybrid functional benchmarks (HSE06:
$E_g = 0.93$~eV)~\cite{hse06}. Without the Hubbard correction, standard PBE severely underestimates the gap ($E_g \approx
	0.4$~eV) and fails to localize the vacancy-induced magnetic moments.

Ion-electron interactions were described using the Projector Augmented Wave (PAW) method~\cite{paw} with a plane-wave kinetic
energy cutoff of $E_{\text{cut}} = 520$~eV. Convergence tests confirmed energy variations $< 1$~meV/atom for $E_{\text{cut}}
	\ge 500$~eV. A vacuum spacing of 20~\AA\ was inserted along the $z$-axis to eliminate spurious interaction between periodic
images.

Vacancy concentrations spanning $3.1\% \le x \le 22.2\%$ were modeled using supercells ranging from $5\times5\times1$ to
$3\times3\times1$. The lattice constant was fixed to the optimized pristine value ($a = 3.40$~\AA) to isolate electronic defect
effects from strain. Internal atomic coordinates were fully relaxed using the conjugate gradient algorithm until residual forces
fell below 0.01~eV/\AA. The electronic self-consistency convergence criterion was set to $10^{-5}$~eV.

Brillouin zone integration was performed using $\Gamma$-centered Monkhorst-Pack grids~\cite{mp} scaled inversely with supercell
size to maintain consistent sampling density: $5\times5\times1$ for the $3\times3$ cell, $4\times4\times1$ for the $4\times4$
cell, and $3\times3\times1$ for the $5\times5$ cell. A denser $12\times12\times1$ grid with Gaussian smearing ($\sigma = 0.05$~eV) was used for density of states (DOS) calculations.

The thermodynamic stability of defect configurations was assessed via the formation energy per formula unit:
\begin{equation}
	E_f = \frac{1}{N}\left(E_{\text{defect}} - m\mu_{\ce{S}} - m'\mu_{\ce{Ti}}\right)
	\label{eq:formation}
\end{equation}
where $m$ and $m'$ are the total number of S and Ti atoms, and $N$ is the number of formula units. The defect formation energy per vacancy was computed as:
\begin{equation}
	E_{df} = \frac{1}{n}\left(E_{\text{defect}} - E_{\text{pristine}} + n\mu_{\ce{S}}\right)
	\label{eq:defect_formation}
\end{equation}
where $n$ is the number of sulfur vacancies and $\mu_{\ce{S}}$ is the chemical potential of sulfur derived from bulk $\alpha$-S. The magnetic ground state was determined by comparing total energies of ferromagnetic (FM) and antiferromagnetic (AFM) spin configurations.

Local magnetic moments ($m_{\ce{Ti}}$) were integrated within Wigner-Seitz spheres ($R_{\text{Ti}} = 1.32$~\AA, $R_{\text{S}} = 1.16$~\AA) as implemented in VASP. Electronic band structures were calculated along the high-symmetry path $\Gamma$--M--K--$\Gamma$ of the hexagonal Brillouin zone.

\subsection{Percolation Analysis of Vacancy Networks}
\label{sub:percolation}

To quantify the geometric connectivity of the vacancy network, we performed a continuum percolation analysis based on
self-consistent charge density distributions. The local charge depletion induced by vacancy formation is defined as:
\begin{equation}
	\Delta\rho(\mathbf{r}) = \rho_{\text{defect}}(\mathbf{r}) - \rho_{\text{pristine}}(\mathbf{r})
	\label{eq:charge_diff}
\end{equation}
where $\rho(\mathbf{r})$ denotes the total electron density evaluated on the DFT Fast Fourier Transform (FFT) grid
($216\times216\times1$ for the $4\times4$ supercell). Since vacancies act as charge sinks, $\Delta\rho$ is predominantly
negative in the defect vicinity. The spatial extent of $\Delta\rho$ directly reflects the range of Ti--Ti magnetic exchange
coupling mediated by the vacancy.

We convert the continuous field into a binary percolation lattice using an occupancy mask:
\begin{equation}
	M(\mathbf{r}) = \begin{cases} 1, & \text{if } \Delta\rho(\mathbf{r}) < \rho_{\text{th}} \\ 0, & \text{otherwise} \end{cases}
	\label{eq:vacancy_mask}
\end{equation}
The threshold $\rho_{\text{th}}$ is set to the 8th percentile of the $\Delta\rho$ distribution. This choice is physically
justified: for a single vacancy in a $4\times4$ supercell, the charge depletion zone at this threshold extends exactly to the
nearest-neighbor Ti sites ($\sim 3.4$~\AA), encompassing $\sim 6$--$8\%$ of the total grid volume~\cite{Freysoldt2014}.
Sensitivity tests confirm that the geometric order parameter ($P_\infty$) is robust to threshold variations within the 5th--15th
percentile range.

Connected clusters were identified using the Hoshen-Kopelman algorithm~\cite{hoshen1976percolation} with 6-connectivity on
the triangular lattice, reflecting the crystal structure of the host material. For each configuration, we computed:
\begin{enumerate*}[label=(\roman*)]
	\item the giant cluster fraction $P_\infty = s_{\max}/\sum_i s_i$, serving as the geometric order parameter;
	\item the total cluster count $N_c$; and
	\item the cluster size distribution $n(s)$.
\end{enumerate*}
At the critical threshold, $n(s)$ was fitted to the power-law form $n(s) \sim s^{-\tau}$ to extract the Fisher exponent
$\tau$~\cite{stauffer1994introduction, saberi2015recent} and verify consistency with the universal 2D percolation value
$\tau_{\text{theory}} = 187/91 \approx 2.05$.

\subsection{Tight-Binding Model for Finite-Size Scaling}
\label{sub:tb}

To validate that the percolation transition persists beyond the constraints of finite DFT supercells, we constructed a
semi-empirical tight-binding (TB) model on a large-scale triangular lattice ($L = 80$, $N = L^2 = 6400$ sites). The Hamiltonian
is:
\begin{equation}
	H = \sum_i \epsilon_i c^\dagger_i c_i - \sum_{\langle i,j \rangle} t\, c^\dagger_i c_j - \sum_{\langle\langle i,j \rangle\rangle} t'\, c^\dagger_i c_j + \text{h.c.}
	\label{eq:tb_hamiltonian}
\end{equation}
where $\langle i,j \rangle$ and $\langle\langle i,j \rangle\rangle$ denote nearest-neighbor (NN) and next-nearest-neighbor (NNN) pairs, respectively. Vacancy sites are assigned $\epsilon_i = 0$ while pristine sites are blocked ($\epsilon_i \to +\infty$). The NN hopping amplitude $t \approx 0.15$~eV was extracted from the bandwidth of the Ti $3d$ impurity band in the DFT DOS, and the NNN amplitude $t' = 0.8t$ was calibrated to reproduce the extended range of defect wavefunctions ($\sim 6$~\AA) observed in the DFT charge density analysis.

Including NNN hopping is essential: DFT charge density analysis reveals that vacancy-induced perturbations extend beyond nearest
neighbors ($\sim 3$--$5$~\AA). This extended-range coupling lowers the percolation threshold from the classical NN-only value
($p_c = 0.50$ for a triangular lattice) to the observed $x_c \approx 0.125$.

For each vacancy concentration, 10 independent disorder realizations were generated. The Hamiltonian was diagonalized using
sparse matrix methods (ARPACK), and the DOS was computed via histogram binning ($\Delta E = 0.01$~eV) and averaged over
realizations. Spatial localization was characterized through the Inverse Participation Ratio (IPR):
\begin{equation}
	\text{IPR}^{(n)} = \sum_{i} \left| \phi_{i}^{(n)} \right|^4
	\label{eq:ipr}
\end{equation}
where $\Psi_n = \sum_i \phi_i^{(n)} |i\rangle$ is the $n$-th eigenstate. Localized states yield $\text{IPR} \sim \mathcal{O}(1)$, while extended Bloch-like states give $\text{IPR} \sim \mathcal{O}(1/M)$ with $M$ the number of active sites.

\section{Results}
\label{sec:results}

We map the full vacancy concentration range ($3\% \le x \le 22\%$) to construct the electronic and geometric phase diagram of
\tis. The data reveal a non-monotonic evolution through three distinct regimes---insulating, half-metallic, and
unstable---governed by the geometric connectivity of the defect network rather than by the vacancy count alone.

\subsection{Thermodynamic Stability and Phase Boundaries}
\label{sub:thermo}

\tabref{tab:fulltab} summarizes the energetics and topology of sulfur vacancies across the concentration range. The formation
energy per formula unit ($E_f$) increases monotonically with vacancy concentration, with vacancies remaining thermodynamically
stable ($E_f < 0$) up to $x = 12.5\%$. Beyond $x \approx 20\%$, positive formation energies ($E_f = +0.11$~eV at $x = 22.2\%$)
signal lattice instability, setting a hard upper bound on the functional doping range.

The defect formation energy per vacancy ($E_{df}/n$) exhibits a physically significant non-monotonicity. It reaches a local
minimum of 8.59~eV at $x = 6.2\%$---substantially lower than the isolated vacancy cost of 9.30~eV---indicating an attractive
interaction that favours vacancy pairing. Above this concentration, repulsive interactions dominate, driving the cost to 9.80~eV
at 12.5\%. This non-monotonic defect interaction energy has implications for vacancy ordering during synthesis: spontaneous
clustering at low concentrations could impede the uniform spatial distribution needed for percolation.

\begin{table}[htpb]
	\centering
	\caption{Energetics and topology of S-vacancies. $E_f$ is the formation energy per formula unit;
		$E_{df}/n$ is the defect formation energy per vacancy; $P_\infty$ is the giant cluster fraction;
		Mean Size is the average cluster size in FFT grid points. The jump from $P_\infty < 5\%$ to
		$P_\infty = 30.4\%$ at 12.5\% marks the percolation transition.}
	\label{tab:fulltab}
	\resizebox{\columnwidth}{!}{%
		\begin{tabular}{ccccccc}
			\toprule
			{Size}     & {Vac (\#)} & {Conc (\%)} & {$E_f$}   & {$E_{df}/n$}  & {$P_\infty$}  & {Mean Size} \\
			           &            &             & (eV/f.u.) & (eV/vac)      & (\%)          & (sites)     \\ \midrule
			$4\times4$ & 1          & 3.1         & $-0.69$   & 9.30          & 4.5           & 8.6         \\
			$4\times4$ & 2          & 6.2         & $-0.63$   & \textbf{8.59} & 3.3           & 11.6        \\
			$5\times5$ & 4          & 8.0         & $-0.48$   & 8.90          & 1.9           & 6.2         \\
			$4\times4$ & 3          & 9.4         & $-0.39$   & 9.21          & 4.0           & 9.0         \\
			$4\times4$ & 4          & 12.5        & $-0.09$   & 9.80          & \textbf{30.4} & 9.6         \\ \midrule
			$3\times3$ & 4          & 22.2        & $+0.11$   & 8.52          & 7.4           & 16.7        \\
			\bottomrule
		\end{tabular}%
	}
\end{table}

\begin{table}[htpb]
	\centering
	\caption{Magnetic properties and spin polarization across the concentration range. Three regimes
		emerge: an insulating ``dead zone'' at $x \approx 6\%$ with isolated defects; a functional
		half-metallic window peaking at $x = 12.5\%$; and a re-entrant collapse of spin polarization
		at $x \approx 22\%$.}
	\label{tab:spinpol}
	\begin{tabular}{cccccc}
		\toprule
		{Conc.} & {System}   & {$m_{\text{Ti}}^{\text{active}}$} & \multicolumn{2}{c}{{$N(E_F)$}} & {Pol.}           \\
		(\%)    & Size       & ($\mu_B$)                         & Up                             & Dn       & (\%)  \\\midrule
		0.0     & $4\times4$ & 0.00                              & 0.00                           & 0.00     & --    \\
		3.1     & $4\times4$ & 0.94                              & 6.42                           & 3.39     & 31    \\
		6.3     & $4\times4$ & 0.98                              & $\sim 0$                       & $\sim 0$ & --    \\
		8.0     & $5\times5$ & 0.99                              & 0.82                           & 0.56     & 19    \\
		{9.4}   & $4\times4$ & 0.95                              & 0.25                           & 0.00     & {100} \\
		{12.5}  & $4\times4$ & {0.77}                            & {38.96}                        & {0.00}   & {100} \\
		22.2    & $3\times3$ & 1.00                              & 9.35                           & 5.51     & 26    \\
		\bottomrule
	\end{tabular}
\end{table}

\subsection{Electronic Structure Evolution}
\label{sub:electronic}

The electronic structure evolves through four qualitatively distinct regimes as the vacancy concentration is increased
(\tabref{tab:spinpol}).

\begin{figure*}[htbp]
	\centering
	\begin{subfigure}{0.24\textwidth}
		\includegraphics[width=\textwidth]{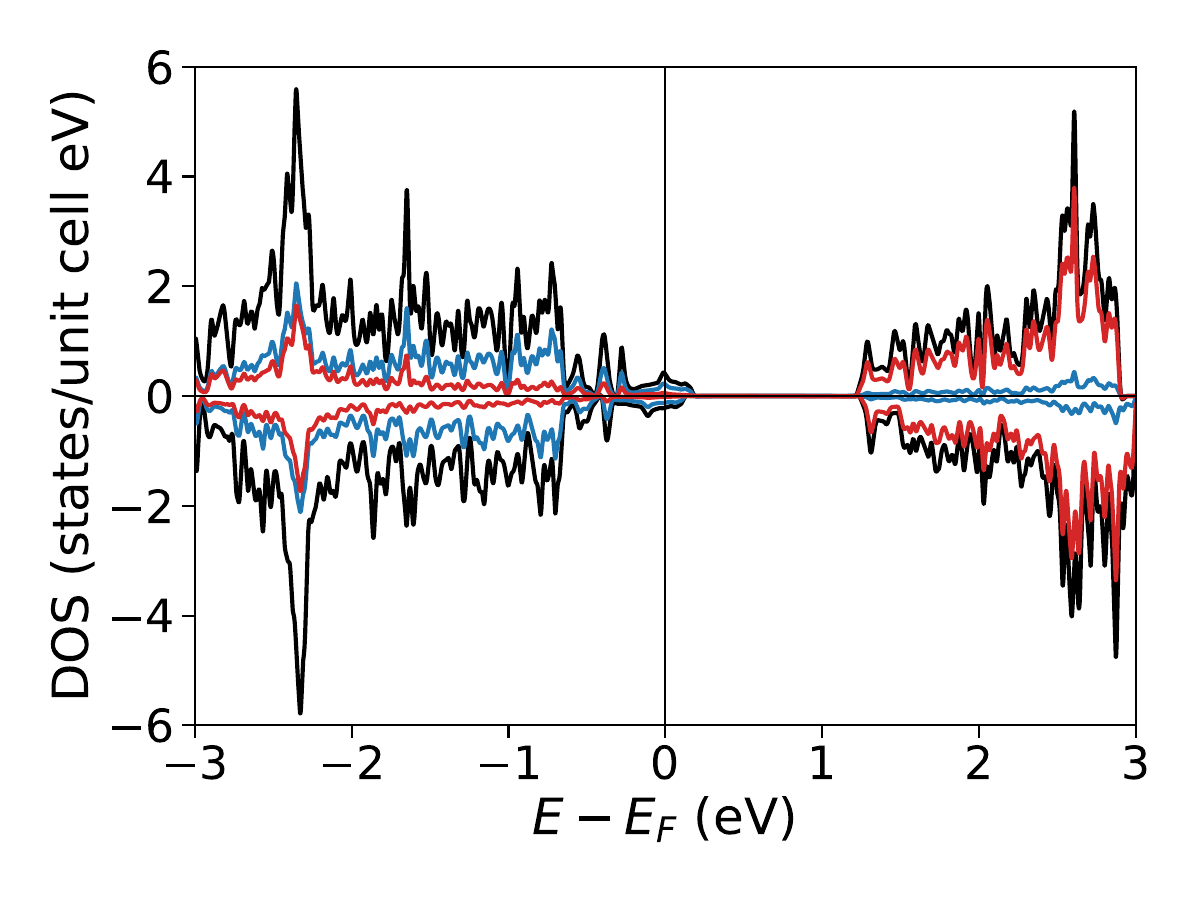}
		\caption{$x$=\stripzero03 \%(\ce{TiS_{\fpeval{2-.02* 03}}})}
		\label{fig:dos_03}
	\end{subfigure}
	\begin{subfigure}{0.24\textwidth}
		\includegraphics[width=\textwidth]{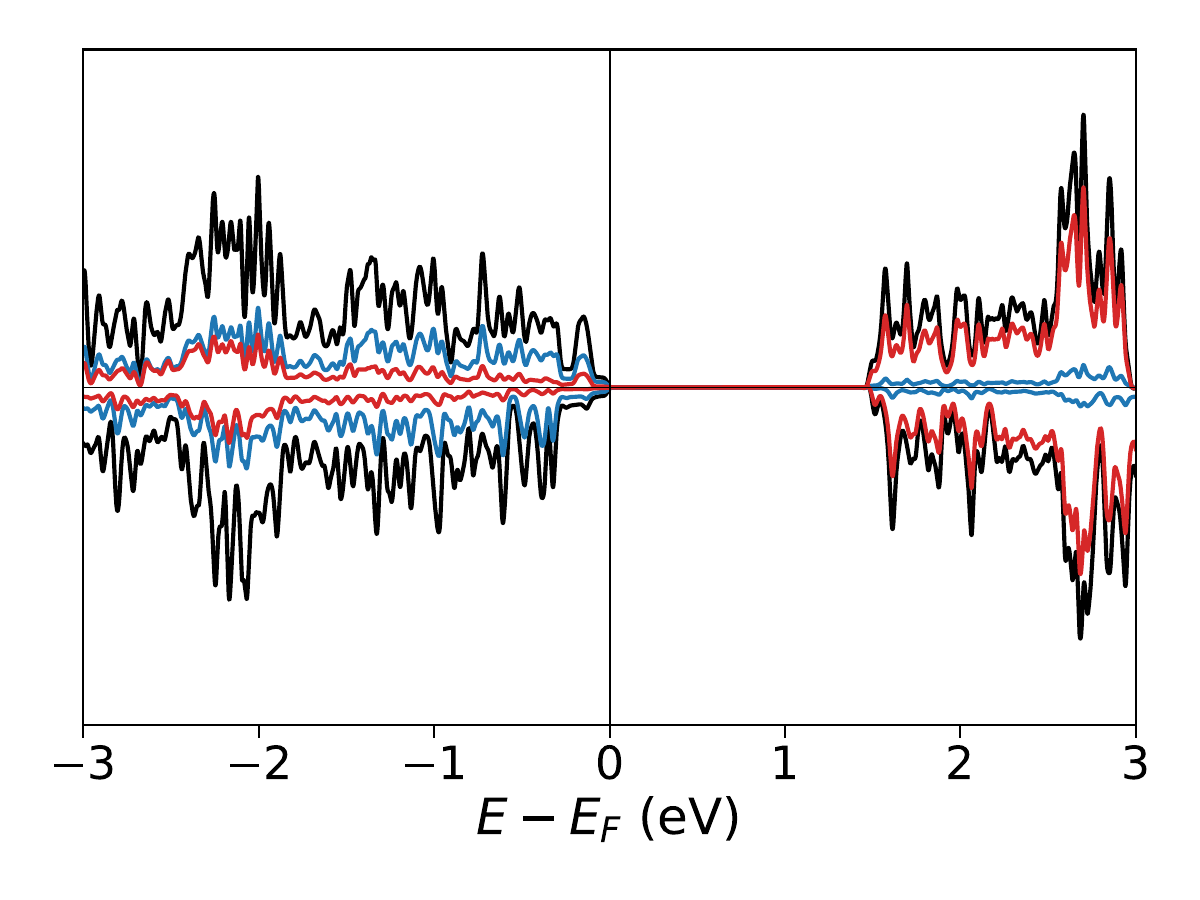}
		\caption{$x$=\stripzero06 \%(\ce{TiS_{\fpeval{2-.02* 06}}})}
		\label{fig:dos_06}
	\end{subfigure}
	\begin{subfigure}{0.24\textwidth}
		\includegraphics[width=\textwidth]{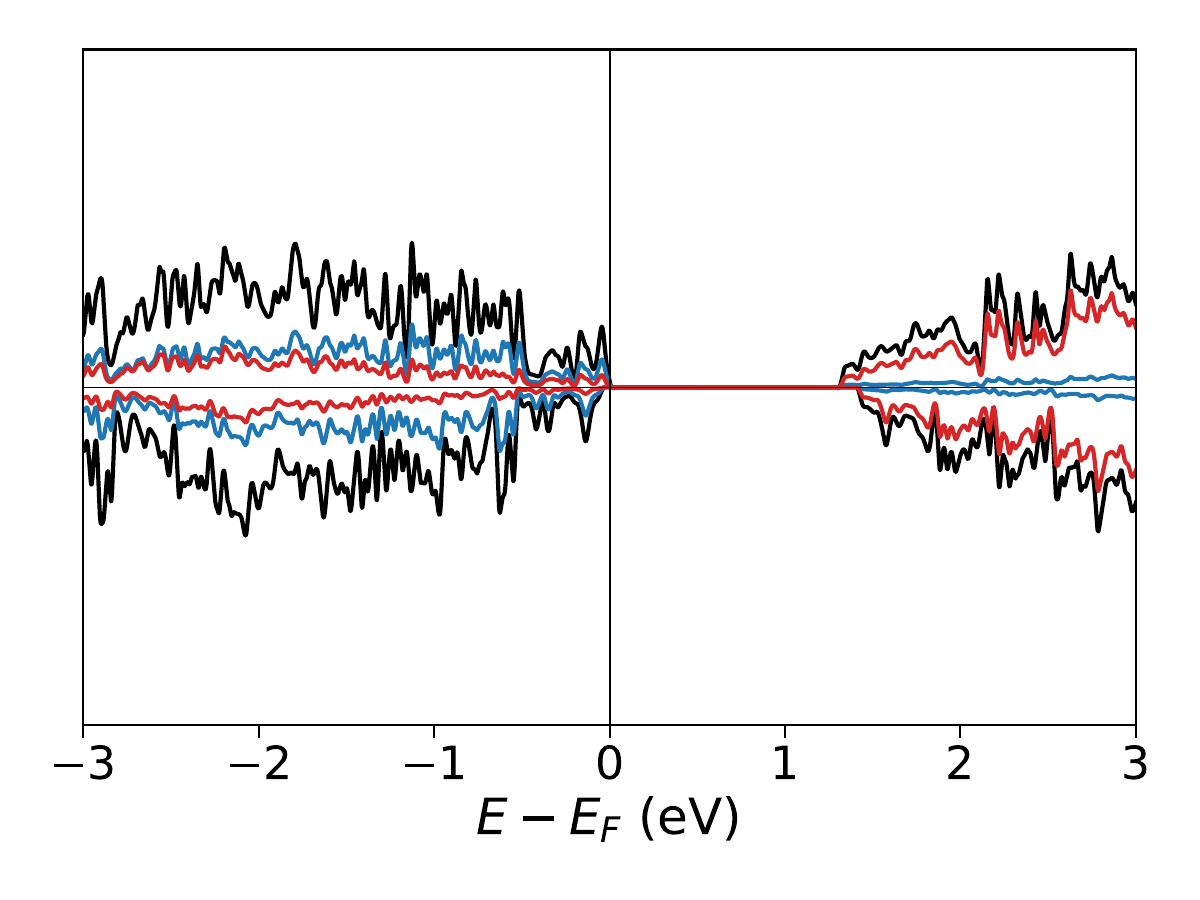}
		\caption{$x$=\stripzero09 \%(\ce{TiS_{\fpeval{2-.02* 09}}})}
		\label{fig:dos_09}
	\end{subfigure}
	\begin{subfigure}{0.24\textwidth}
		\includegraphics[width=\textwidth]{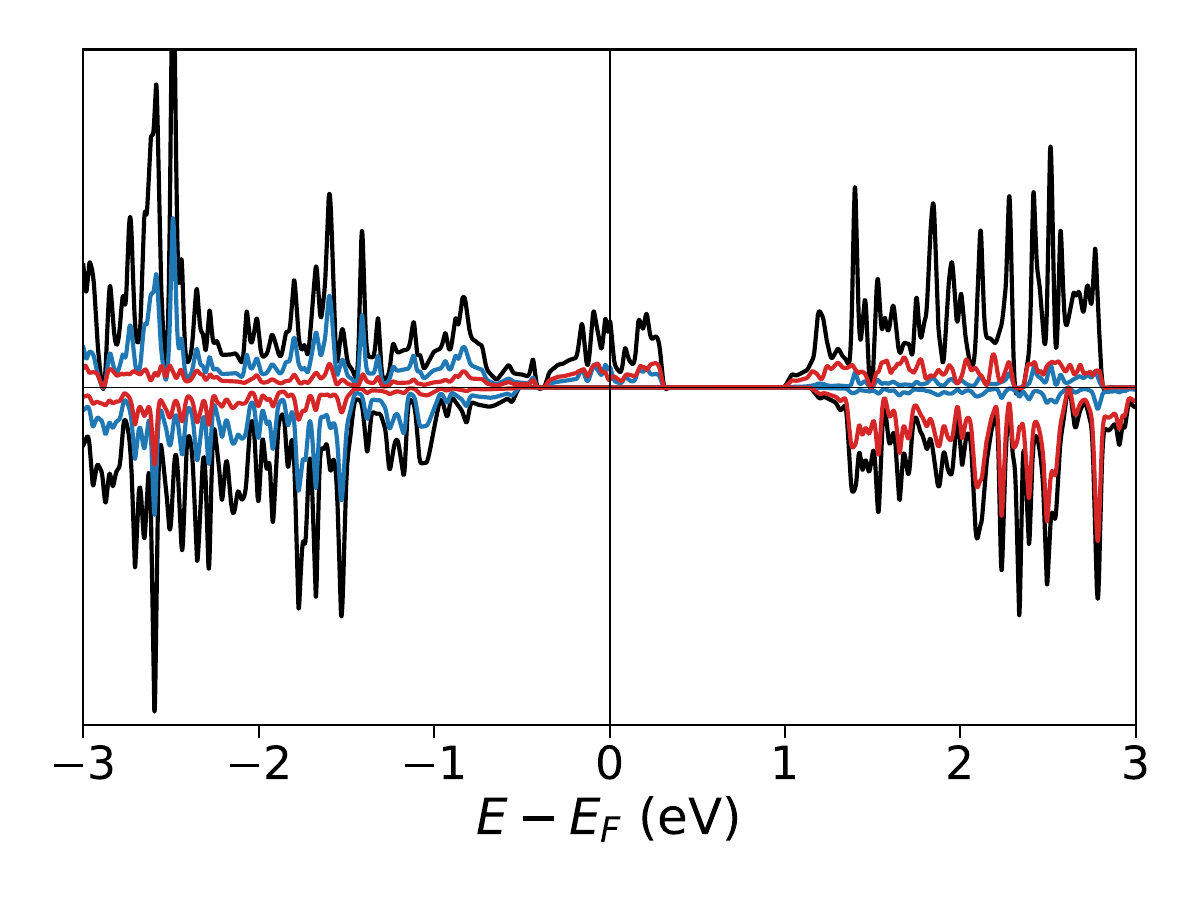}
		\caption{$x$=\stripzero12 \%(\ce{TiS_{\fpeval{2-.02* 12}}})}
		\label{fig:dos_12}
	\end{subfigure}
	\caption{Spin-resolved density of states across the concentration range.
		(\subref{fig:dos_03}) Dilute limit: discrete defect states deep in the gap.
		(\subref{fig:dos_06}) Dead zone: states broaden but remain distinct from $E_F$.
		(\subref{fig:dos_09}) Incipient transition: the majority-spin tail grazes $E_F$.
		(\subref{fig:dos_12}) Percolation threshold: a massive majority-spin peak is pinned at $E_F$
		with a clean minority-spin gap, establishing full half-metallicity.}
	\label{fig:dos}
\end{figure*}

\begin{figure*}[htbp]
	\centering
	\begin{subfigure}{0.24\textwidth}
		\includegraphics[width=\textwidth]{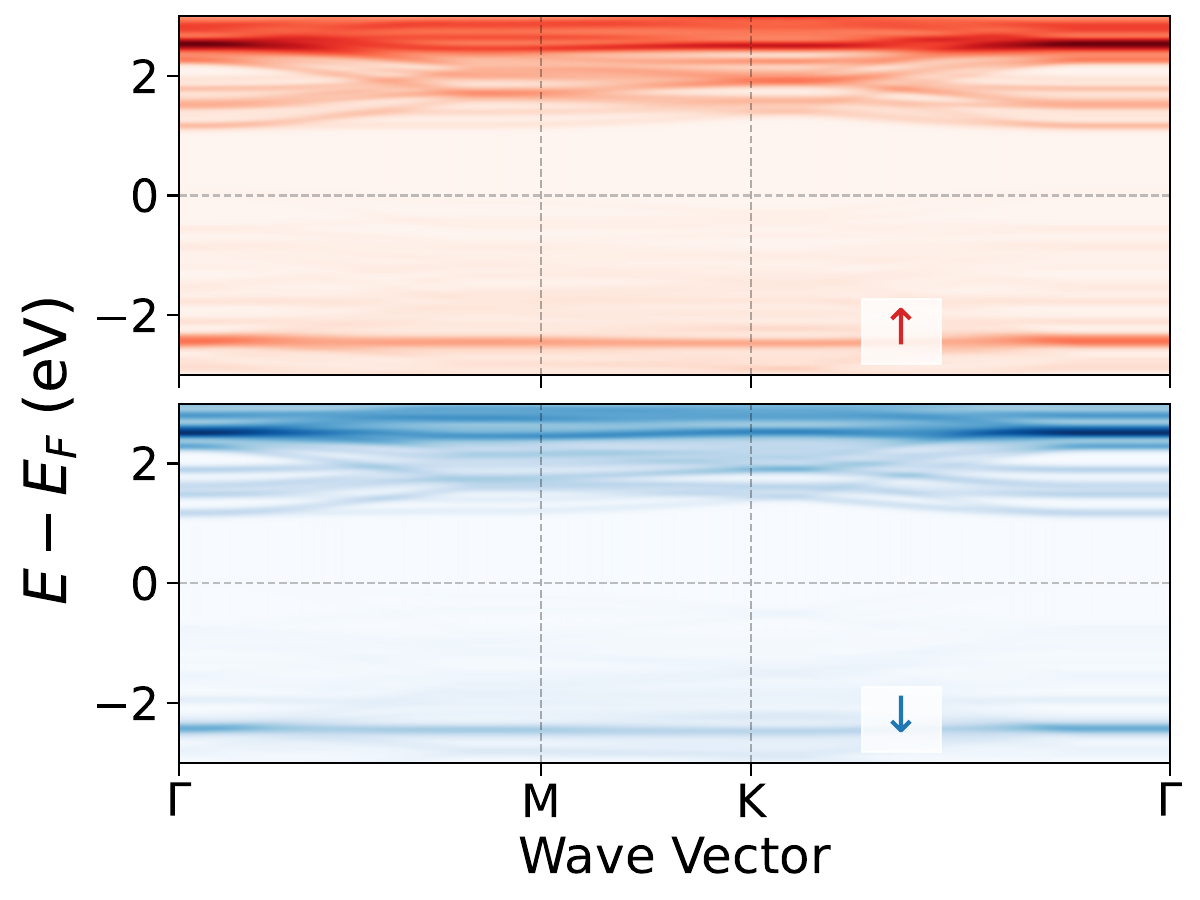}
		\caption{$x$=\stripzero03 \%(\ce{TiS_{\fpeval{2-.02* 03}}})}
		\label{fig:band_03}
	\end{subfigure}
	\begin{subfigure}{0.24\textwidth}
		\includegraphics[width=\textwidth]{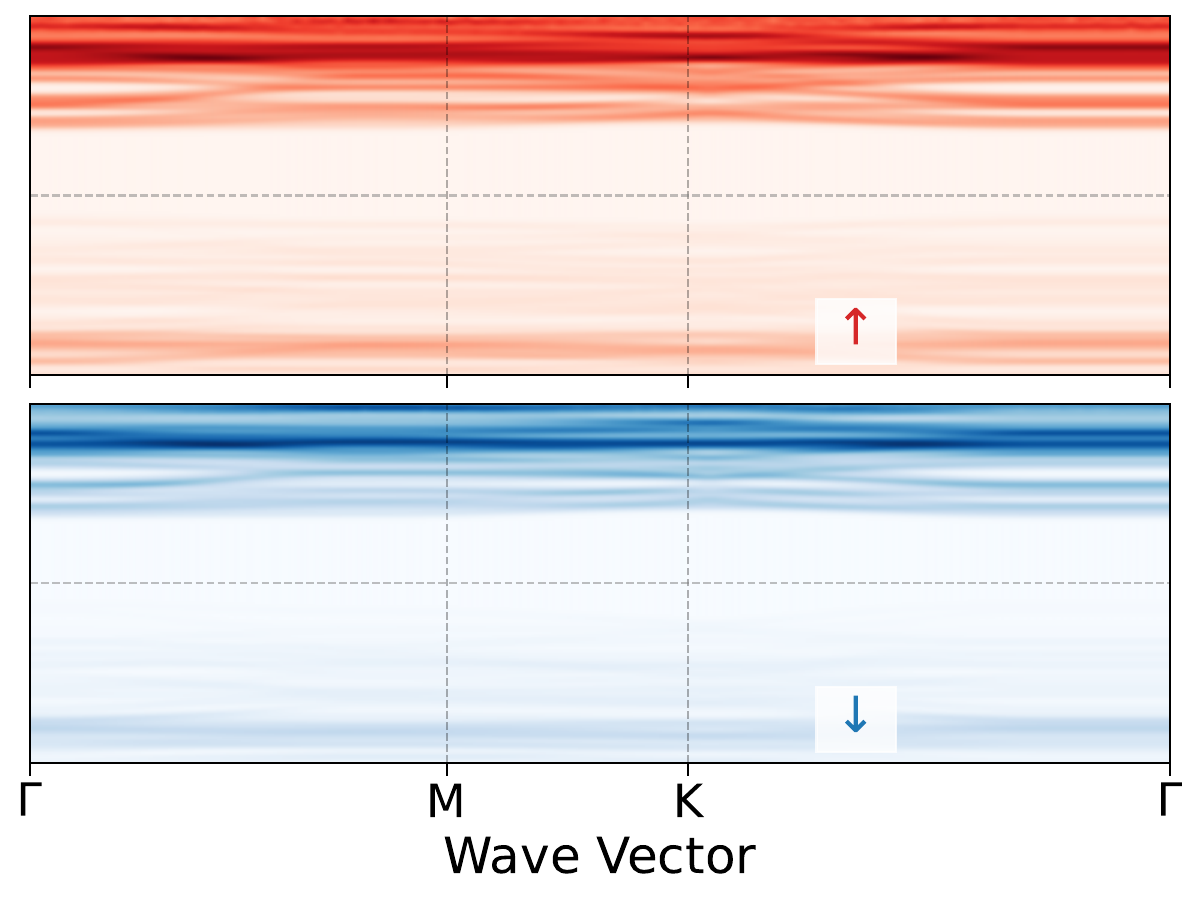}
		\caption{$x$=\stripzero06 \%(\ce{TiS_{\fpeval{2-.02* 06}}})}
		\label{fig:band_06}
	\end{subfigure}
	\begin{subfigure}{0.24\textwidth}
		\includegraphics[width=\textwidth]{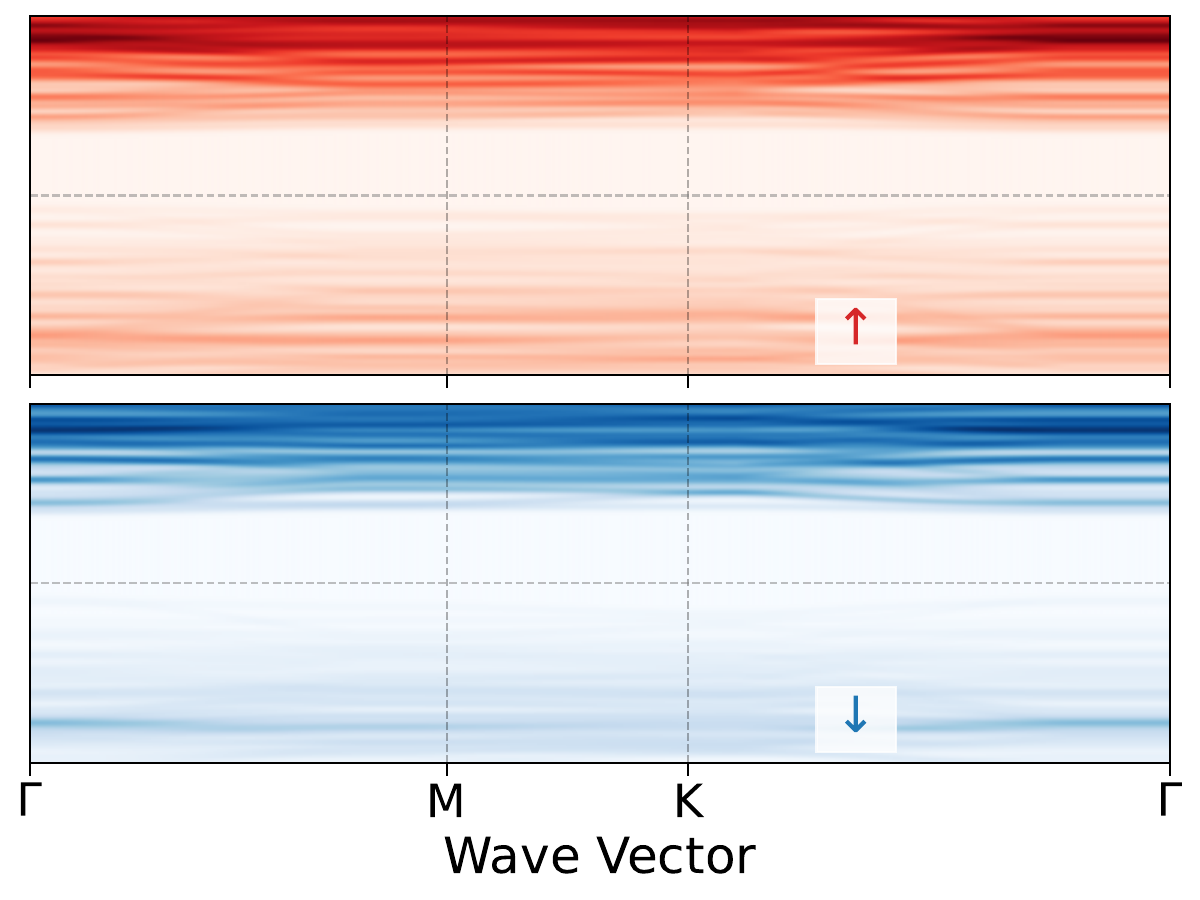}
		\caption{$x$=\stripzero09 \%(\ce{TiS_{\fpeval{2-.02* 09}}})}
		\label{fig:band_09}
	\end{subfigure}
	\begin{subfigure}{0.24\textwidth}
		\includegraphics[width=\textwidth]{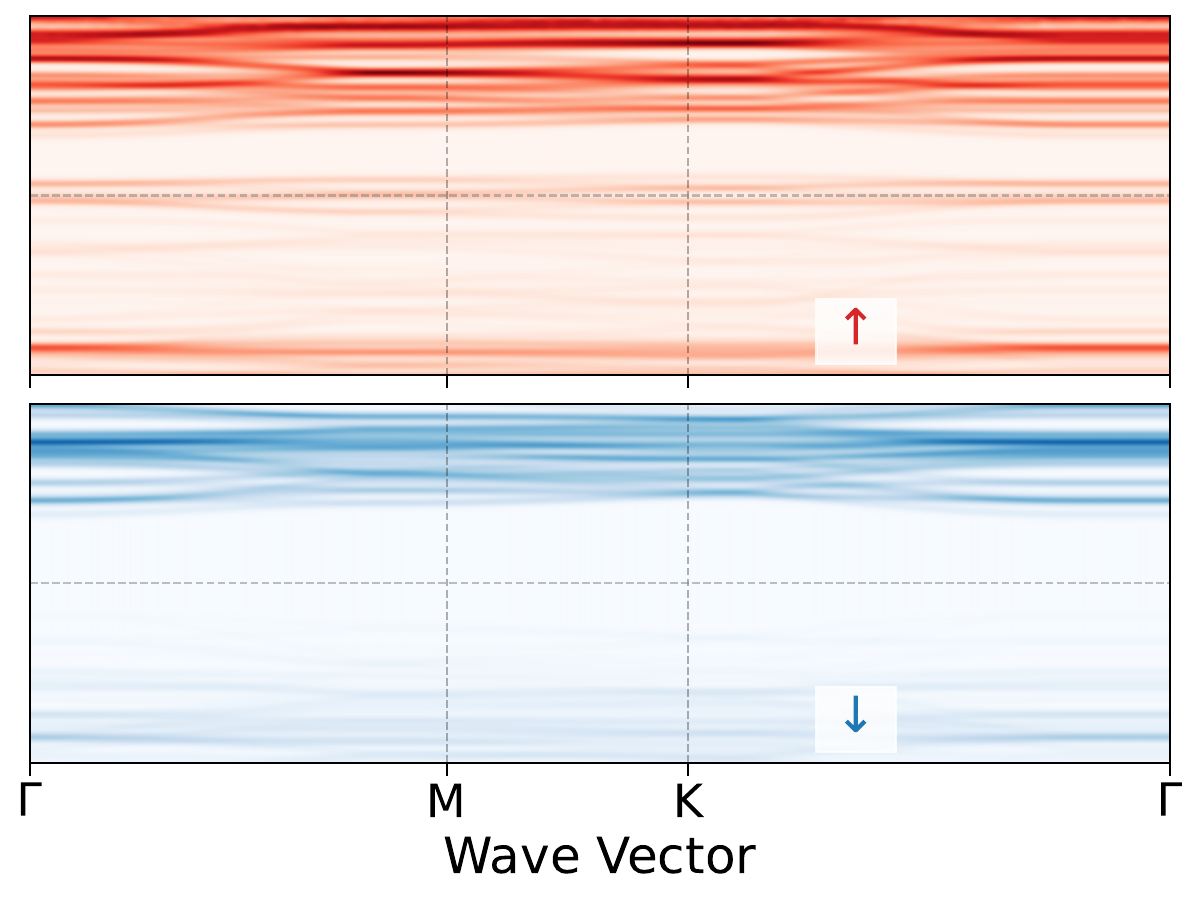}
		\caption{$x$=\stripzero12 \%(\ce{TiS_{\fpeval{2-.02* 12}}})}
		\label{fig:band_12}
	\end{subfigure}
	\caption{Band structure evolution from localized moments to itinerant transport.
		(\subref{fig:band_03}--\subref{fig:band_09}) Below threshold, defect-induced levels are flat and
		dispersionless ($W < 0.1$~eV), indicating negligible wavefunction overlap.
		(\subref{fig:band_12}) At $x = 12\%$, these levels hybridize into a 1.5~eV-wide dispersive band
		that crosses $E_F$---the spectroscopic fingerprint of the percolation transition.}
	\label{fig:bands}
\end{figure*}

\begin{figure*}[htbp]
	\centering
	\begin{subfigure}{0.24\textwidth}
		\includegraphics[width=\textwidth]{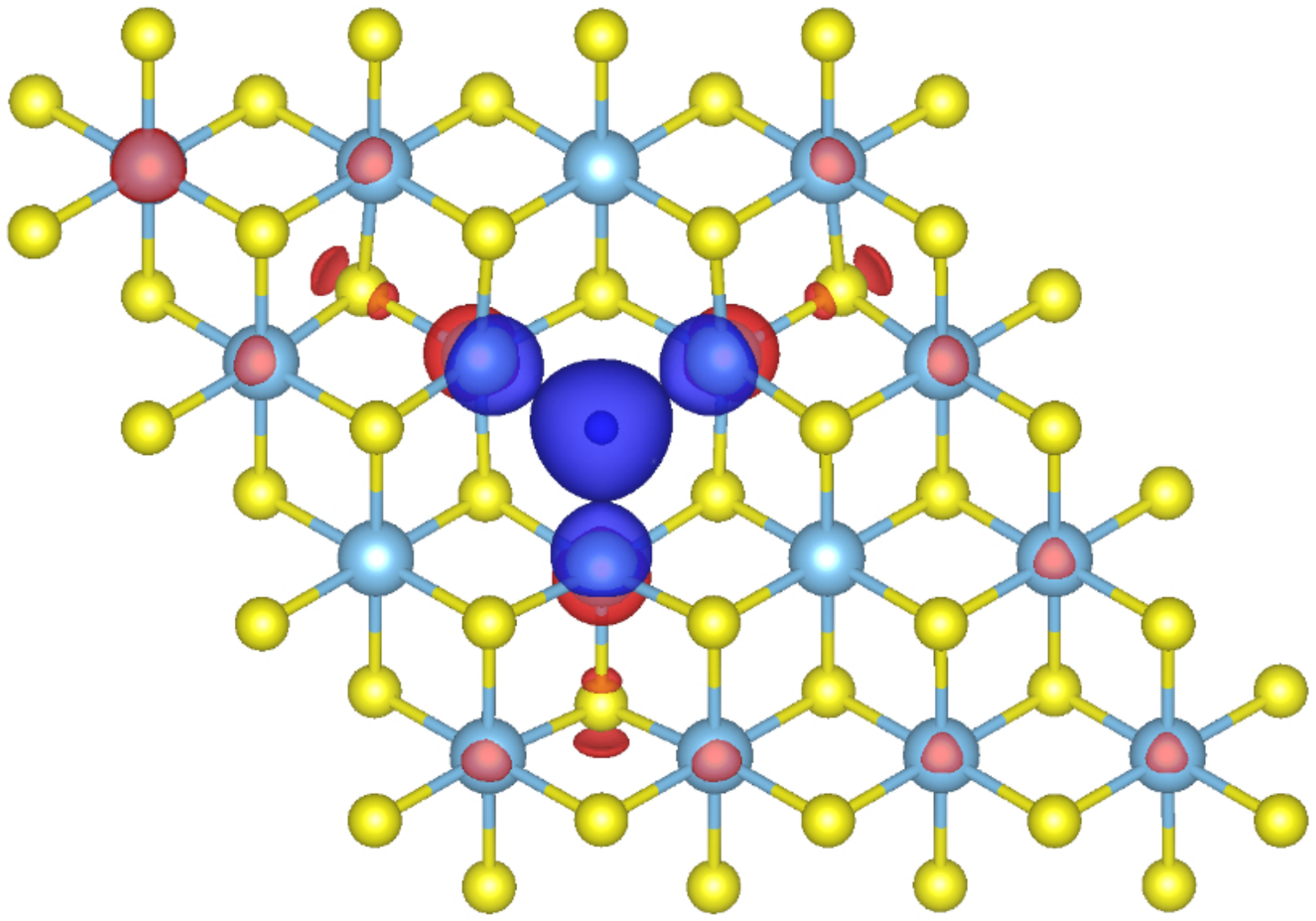}
		\caption{$x$=\stripzero03 \%(\ce{TiS_{\fpeval{2-.02* 03}}})}
		\label{fig:chg_03}
	\end{subfigure}
	\begin{subfigure}{0.24\textwidth}
		\includegraphics[width=\textwidth]{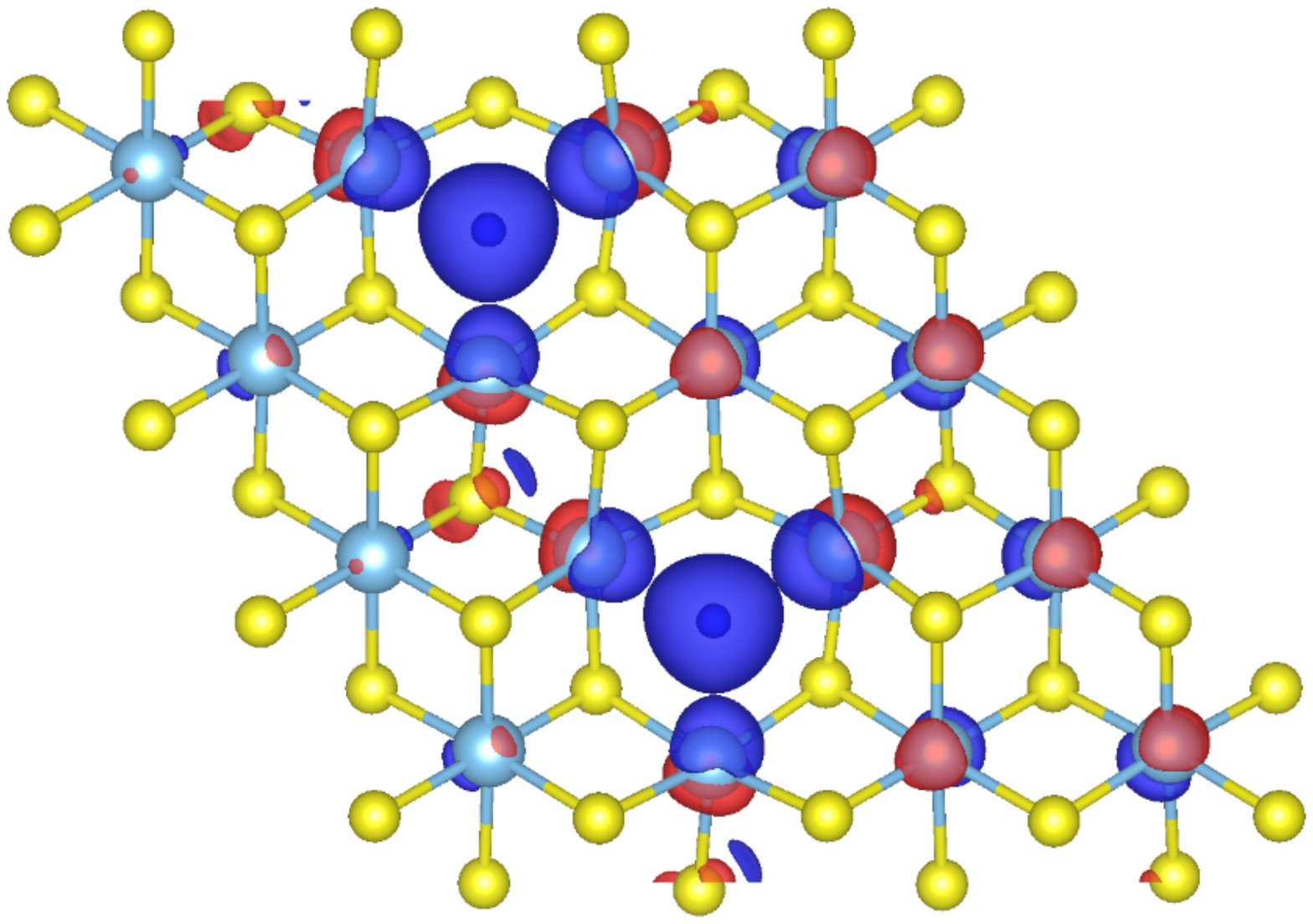}
		\caption{$x$=\stripzero06 \%(\ce{TiS_{\fpeval{2-.02* 06}}})}
		\label{fig:chg_06}
	\end{subfigure}
	\begin{subfigure}{0.24\textwidth}
		\includegraphics[width=\textwidth]{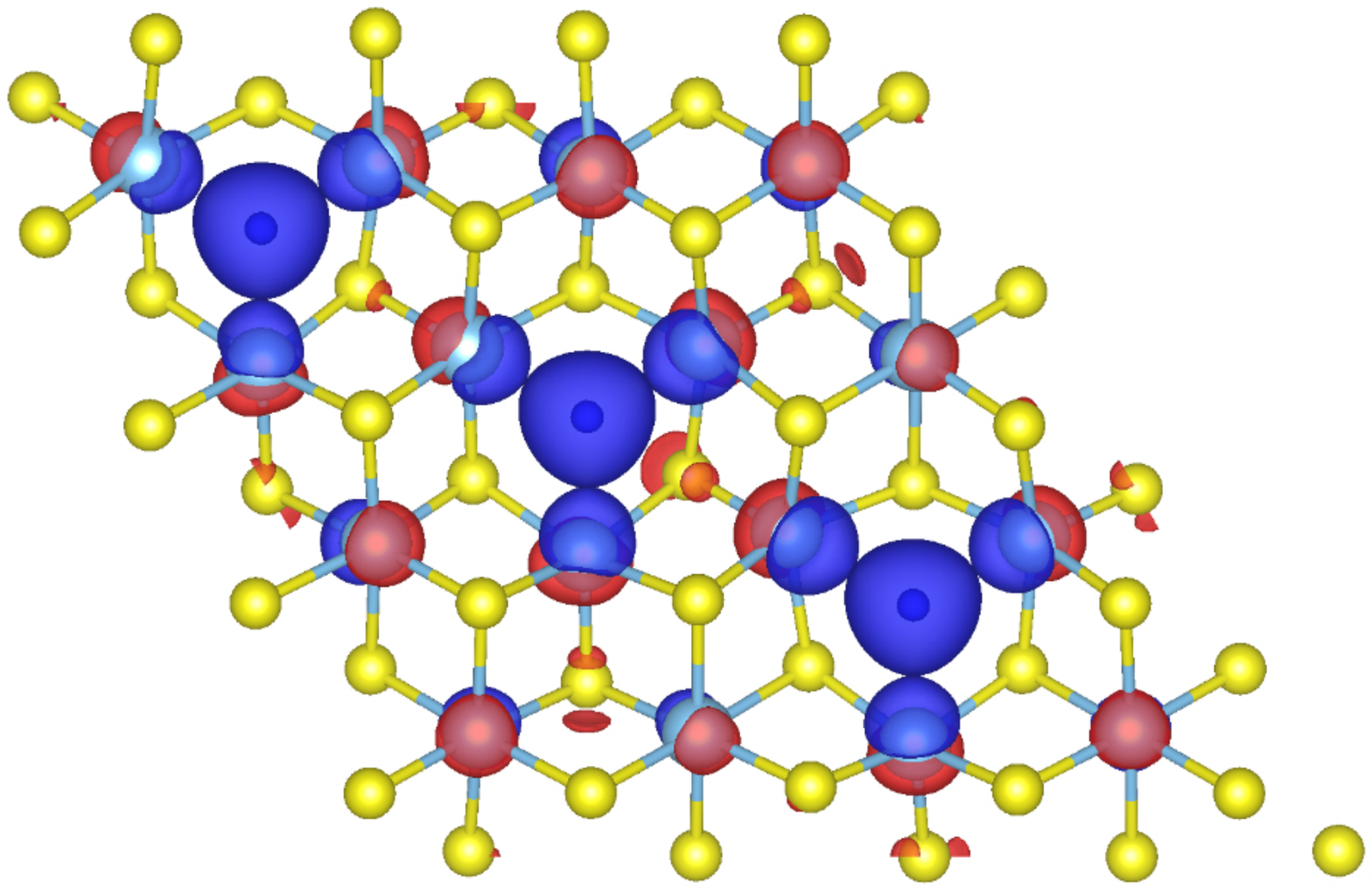}
		\caption{$x$=\stripzero09 \%(\ce{TiS_{\fpeval{2-.02* 09}}})}
		\label{fig:chg_09}
	\end{subfigure}
	\begin{subfigure}{0.24\textwidth}
		\includegraphics[width=\textwidth]{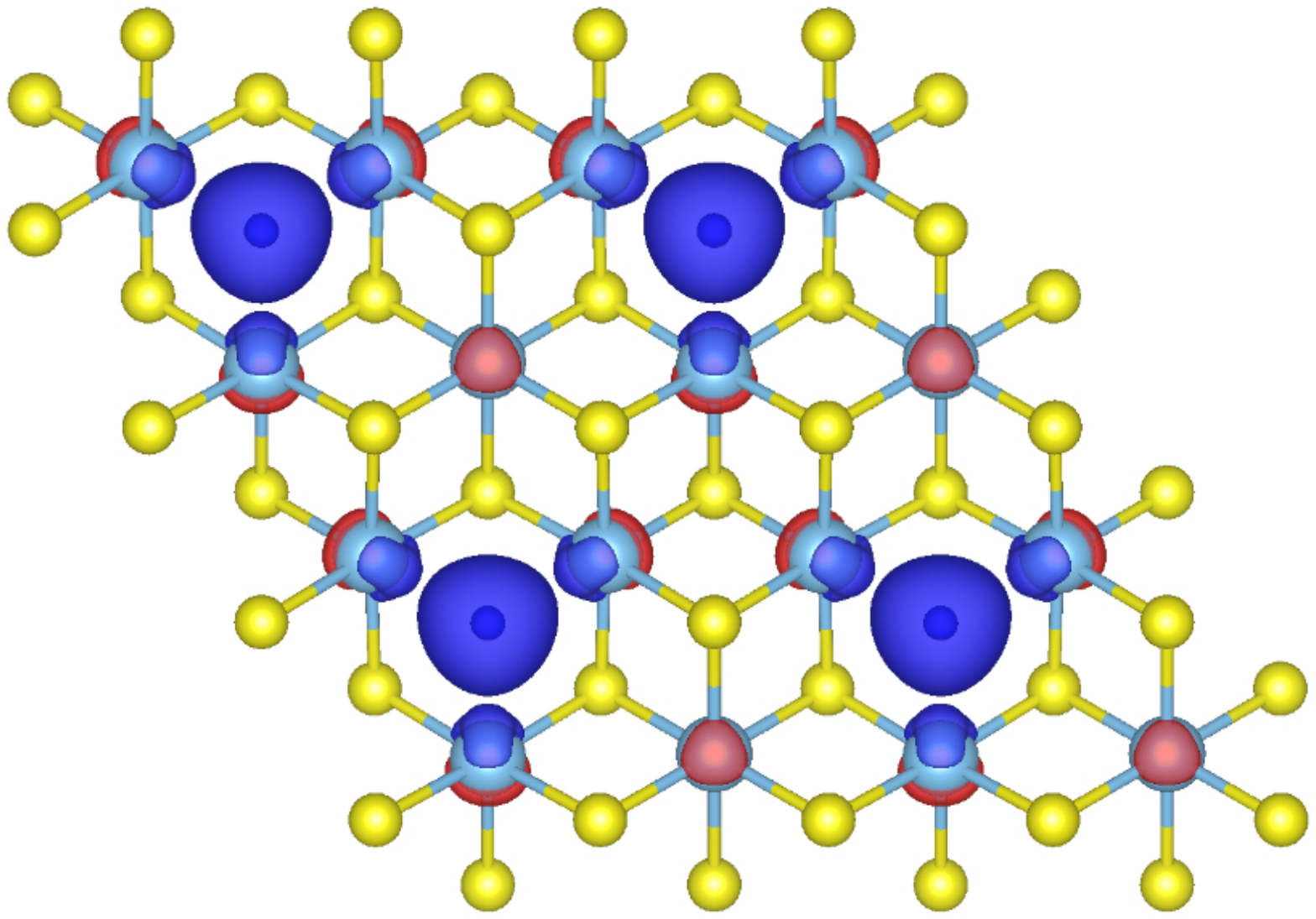}
		\caption{$x$=\stripzero12 \%(\ce{TiS_{\fpeval{2-.02* 12}}})}
		\label{fig:chg_12}
	\end{subfigure}
	\caption{Charge density difference ($\Delta\rho = \rho_{\text{defect}} - \rho_{\text{pristine}}$)
		visualizing the percolation mechanism.
		(\subref{fig:chg_03}) Dilute: the perturbation is confined to nearest-neighbour Ti atoms.
		(\subref{fig:chg_06}--\subref{fig:chg_09}) Intermediate: charge clouds remain spatially disjoint.
		(\subref{fig:chg_12}) Threshold: defect orbitals coalesce into a continuous network spanning the
		supercell.}
	\label{fig:chgs}
\end{figure*}

\textit{Dilute limit ($x = 3.1\%$).}---In the dilute limit, each sulfur vacancy creates a pair of spin-split Ti $3d$ impurity
states deep within the pristine band gap (\figref{fig:dos_03}). The corresponding band structure (\figref{fig:band_03})
reveals flat, dispersionless levels with bandwidth $W < 0.1$~eV, confirming that carriers are Anderson-localized within isolated
magnetic islands. The charge density difference (\figref{fig:chg_03}) shows strict spatial confinement of the electronic
perturbation to the nearest-neighbour Ti atoms ($r_{\text{eff}} \approx 3.5$~\AA). Each vacancy generates a robust local moment
of $0.94~\mu_B$.

\textit{Dead zone ($x = 6.3\%$).}---Doubling the defect density produces a counter-intuitive result: the DOS at $E_F$
\emph{decreases} to negligible values (\figref{fig:dos_06}), and the system reverts to an insulator despite having twice as many
magnetic centers. This ``dead zone'' arises because the two vacancies in the $4\times4$ supercell adopt a configuration in which
their charge-depletion clouds remain disjoint (\figref{fig:chg_06}). The attractive defect interaction (minimum $E_{df}/n$ at
this concentration, \tabref{tab:fulltab}) favours vacancy pairing, but the paired configuration produces an isolated dimer rather
than an extended network. The local moments ($0.98~\mu_B$) are individually robust but collectively impotent: without a
connected exchange pathway, they cannot sustain itinerant transport.

\textit{Incipient transition ($x = 9.4\%$).}---At three vacancies per $4\times4$ cell, the majority-spin impurity states graze
$E_F$ for the first time, yielding 100\% spin polarization but with a vanishingly small DOS of only 0.25~states/eV
(\figref{fig:dos_09}). The band structure (\figref{fig:band_09}) remains essentially flat, and the local moment ($0.95~\mu_B$)
is barely reduced from the dilute value. This regime represents a fragile, incipient half-metallicity in which a handful of
majority-spin states touch the Fermi level without forming a robust conducting channel. The geometric order parameter confirms the
diagnosis: $P_\infty = 4.0\%$, well below the percolation threshold.

\textit{Percolation threshold ($x = 12.5\%$).}---The transition to robust half-metallicity is abrupt.
The DOS (\figref{fig:dos_12}) reveals the hallmark of a half-metal: the Fermi level intersects a continuum of majority-spin
states ($N_\uparrow(E_F) \approx 39$~states/eV) while the minority-spin channel retains a clean insulating gap of
$\Delta_\downarrow = 1.0$~eV, resulting in perfect 100\% spin polarization. The definitive signature of this transition is
captured in the band structure (\figref{fig:band_12}): the majority-spin impurity levels undergo a dramatic 15-fold broadening,
evolving from flat levels ($W < 0.1$~eV) into a dispersive band with $W \approx 1.5$~eV that crosses $E_F$. This asymmetric
delocalization---majority-spin dispersive, minority-spin unchanged---provides strong evidence for a carrier-mediated
double-exchange mechanism, in which itinerant electrons hop between ferromagnetically aligned sites without spin-flip scattering.
The charge density difference (\figref{fig:chg_12}) reveals the geometric origin: defect orbitals coalesce into a continuous,
system-spanning network. The simultaneous reduction of the local moment to $0.77~\mu_B$ confirms that spectral weight has been
redistributed from localized to itinerant states.

\textit{Geometric jamming ($x = 22.2\%$).}---At high vacancy densities, the system exits the functional window. Despite
retaining large local moments ($1.0~\mu_B$), the spin polarization collapses to 26\% and the formation energy becomes positive
($E_f = +0.11$~eV). The geometric order parameter drops to $P_\infty = 7.4\%$, despite the high local defect density. This
apparent contradiction---high vacancy count but low connectivity---reflects a transition from extended percolation to compact
clustering: vacancies coalesce into dense, isolated droplets rather than forming an extended network, simultaneously destroying
both thermodynamic stability and spin-polarized transport.

\subsection{Orbital Origin of the Local Magnetic Moment}
\label{sub:orbital}

To elucidate the microscopic mechanism that protects the local moment against lattice relaxation---a key distinction between
\ptis~and the Group-VI TMDCs---we performed site-resolved orbital decomposition of the electronic DOS.

\begin{figure*}[htpb]
	\centering
	\begin{subfigure}[b]{.3\textwidth}
		\includegraphics[width=\textwidth]{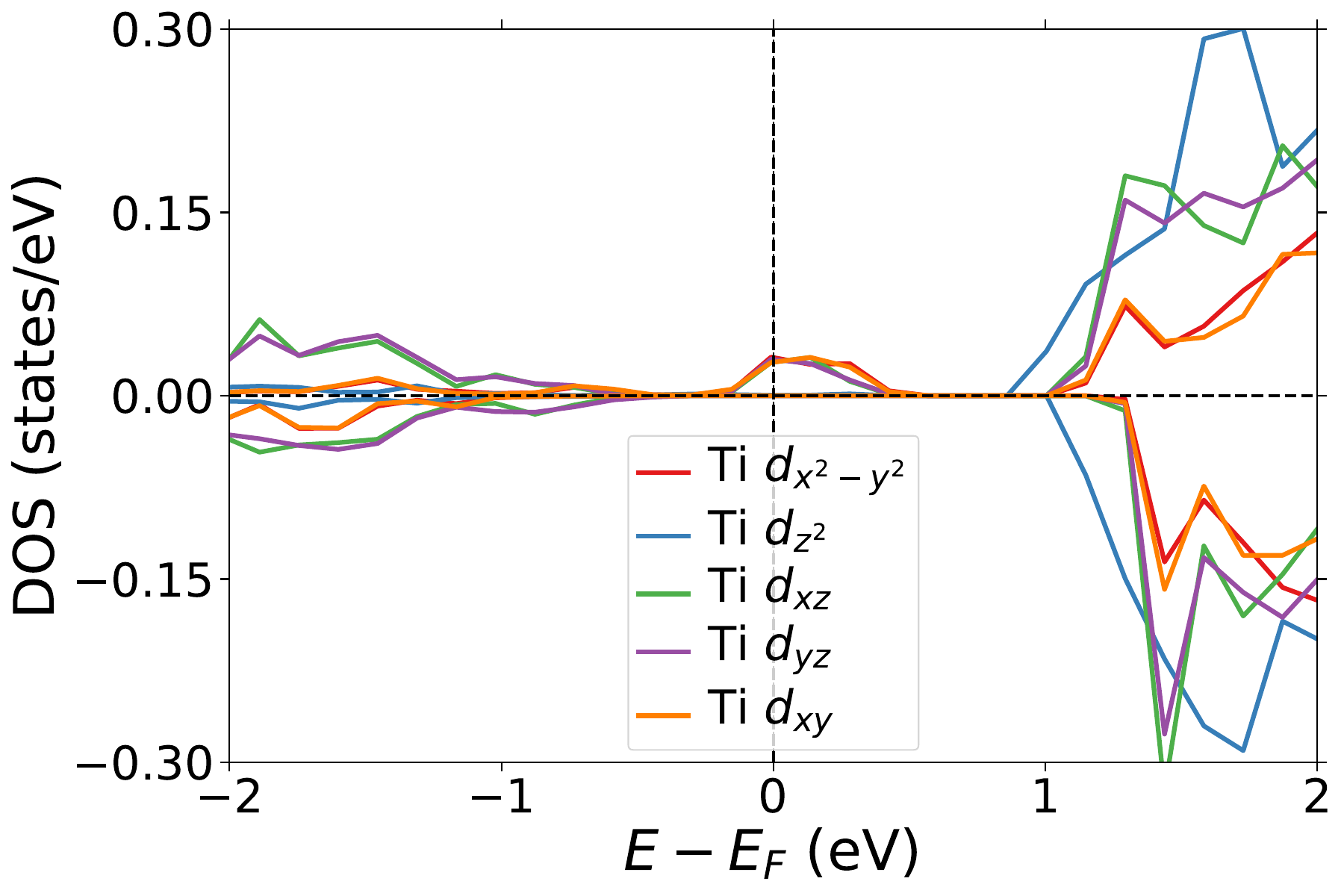}
		\caption{Reference Ti: octahedral $O_h$ field}
		\label{fig:orbital-far}
	\end{subfigure}
	\begin{subfigure}[b]{.3\textwidth}
		\includegraphics[width=\textwidth]{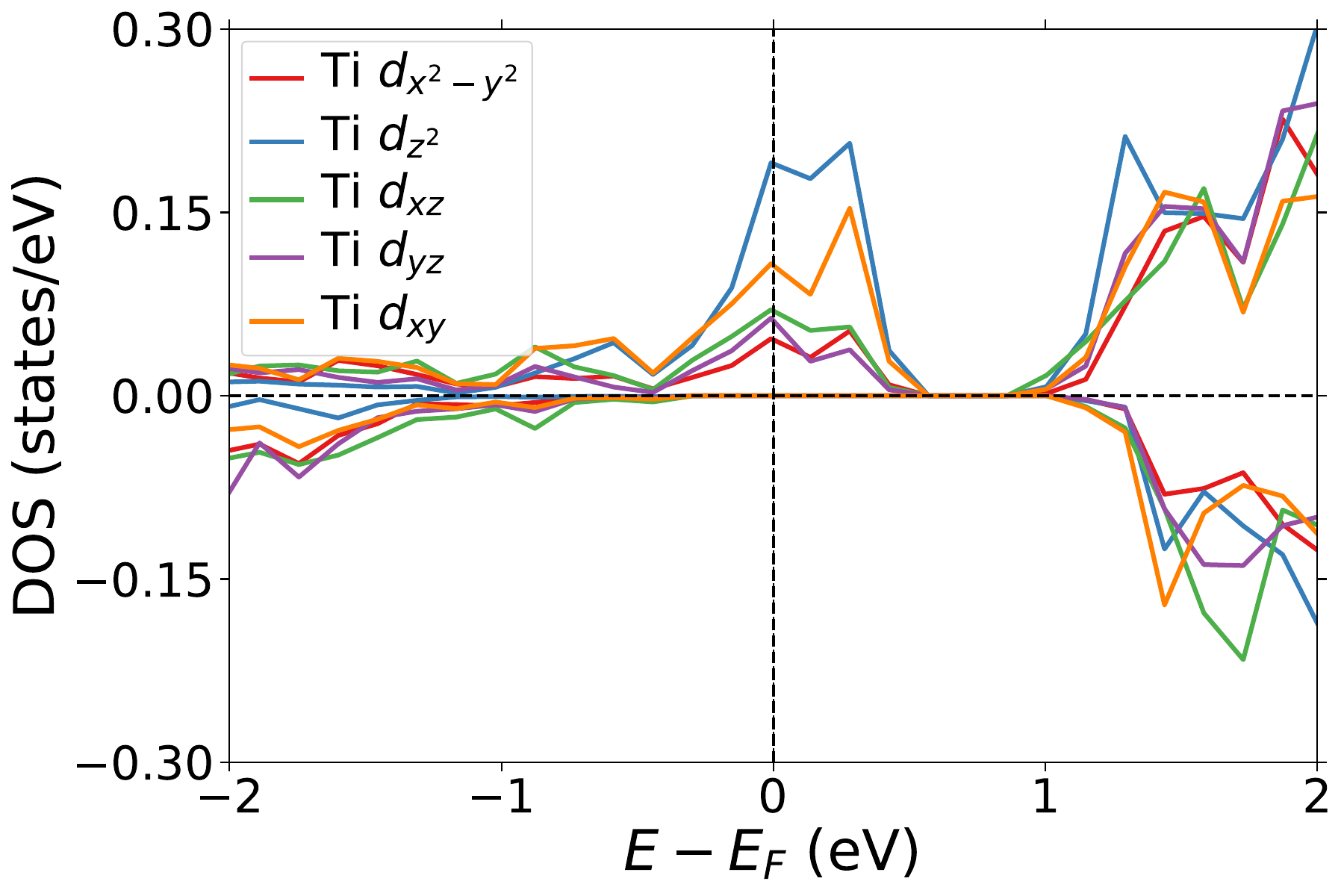}
		\caption{Active Ti: reduced $C_{4v}$ field}
		\label{fig:orbital-near}
	\end{subfigure}
	\begin{subfigure}[b]{.3\textwidth}
		\includegraphics[width=\textwidth]{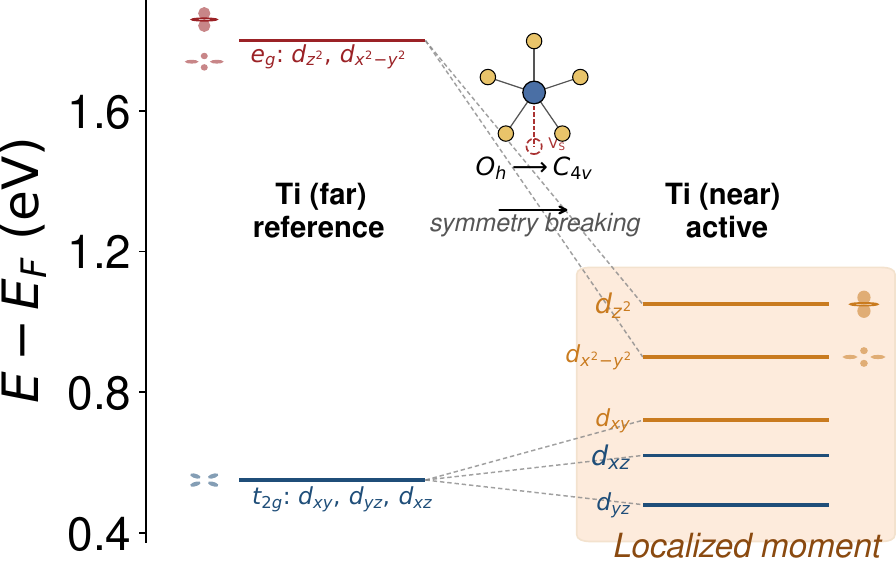}
		\caption{Crystal-field splitting schematic}
		\label{fig:orbital-cfs}
	\end{subfigure}
	\caption{Orbital origin of the local magnetic moment.
	(\subref{fig:orbital-far}) Projected DOS for a bulk-like Ti atom, showing the characteristic $O_h$
	splitting into $t_{2g}$ (valence) and $e_g$ (conduction) sub-bands.
	(\subref{fig:orbital-near}) Projected DOS for the vacancy-adjacent Ti atom. The reduced $C_{4v}$
	symmetry collapses the $e_g$ manifold, producing a mid-gap impurity state with $d_{z^2}/d_{x^2-y^2}$
	character.
	(\subref{fig:orbital-cfs}) Schematic energy-level diagram illustrating the mechanism: the lost
	anion reduces repulsion on the $d_{z^2}$ orbital, stabilizing it within the gap.}
	\label{fig:orbitals}
\end{figure*}

For a reference Ti atom far from the defect (\figref{fig:orbital-far}), the $3d$ manifold exhibits the characteristic octahedral
($O_h$) crystal-field splitting: non-bonding $t_{2g}$ states ($d_{xy}$, $d_{yz}$, $d_{xz}$) form the valence band edge, while
the $e_g$ doublet ($d_{z^2}$, $d_{x^2-y^2}$) lies at higher energy in the conduction band, pushed up by Coulomb repulsion with
the surrounding sulfur $p$-orbitals. The Fermi level resides within a clean gap.

The removal of a sulfur neighbour (\figref{fig:orbital-near}) dramatically reconstructs this electronic landscape. The local
symmetry is reduced from $O_h$ to $C_{4v}$, with two consequences that are central to our story. First, the $e_g$ degeneracy is
lifted. Second, the $d_{z^2}$ orbital---whose lobe is directed toward the now-vacant anion site---experiences a sharp reduction
in Coulomb repulsion. As a result, this orbital detaches from the conduction band and collapses into the band gap, undergoing an
energy shift of approximately 1.0~eV (\figref{fig:orbital-cfs}). The resulting narrow impurity state, with mixed
$d_{z^2}/d_{x^2-y^2}$ character, is spatially localized on the three Ti atoms surrounding the vacancy and subject to strict
on-site occupancy limits. The half-filled impurity level hosts the unpaired electron, generating the localized moment
($\sim 1.0~\mu_B$) that serves as the elementary building block of the percolating ferromagnetic network.

This orbital-selective mechanism explains why \ptis~sustains magnetism while most Group-VI TMDCs do not. In $2H$-\ce{MoS2} and
\ce{WS2}, the larger Mo/W $4d$/$5d$ orbitals undergo substantial inward relaxation around the vacancy, reconstructing the
local bonding environment and quenching the moment~\cite{Gaikwad2025, Hossen2024}. The smaller Ti $3d$ orbitals in the $1T$
octahedral framework instead accommodate charge redistribution without spin quenching, paralleling the vacancy-induced
magnetism observed in phase-engineered 1T'-\ce{MoS2}~\cite{Cai2015}.

\subsection{Magnetic Ground State: Supercell-Size Dependence}
\label{sub:magnetic}

The magnetic coupling mechanism provides direct evidence for the percolation-driven nature of the transition through a striking
dependence on the spatial extent of the vacancy network.

In the $2\times2$ supercell at $x = 12.5\%$, the percolation network is artificially truncated by the periodic boundaries.
Despite being at the critical concentration, this confinement forces the system into a localized antiferromagnetic (AFM) ground
state, with an energy difference $\Delta E_{\text{AFM-FM}} = -0.65$~eV per vacancy favouring the N\'eel configuration. The
physics is that of nearest-neighbour superexchange between a spatially constrained vacancy pair: without a spanning cluster, the
kinetic energy gain from carrier delocalization is unavailable, and localized antiferromagnetic coupling prevails.

Restoring the long-range connectivity in $4\times4$ supercells triggers a fundamental reversal. The ferromagnetic state becomes
the robust ground state, and AFM initializations spontaneously relax to FM order during the self-consistency cycle. The
N\'eel-AFM configuration becomes \emph{dynamically unstable}---it fails to converge entirely---signalling that the
antiferromagnetic manifold is no longer a stationary point on the potential energy surface. The ferromagnetic ground state is
therefore not merely energetically preferred, but \emph{mandated} by the itinerant topology of the spanning cluster: the kinetic
energy gain from carrier delocalization across the infinite network overwhelms the local superexchange cost.

This AFM$\to$FM reversal at identical concentration but different supercell sizes cannot be explained by concentration alone. It
constitutes direct evidence for the percolation mechanism: the $2\times2$ cell is simply too small to host a spanning cluster,
suppressing the double-exchange pathway that stabilizes itinerant ferromagnetism in the $4\times4$ geometry.

\subsection{Percolation Analysis and Phase Synchronization}
\label{sub:percolation_results}

\begin{figure*}[htbp]
	\centering
	\begin{subfigure}{0.32\textwidth}
		\includegraphics[width=\textwidth]{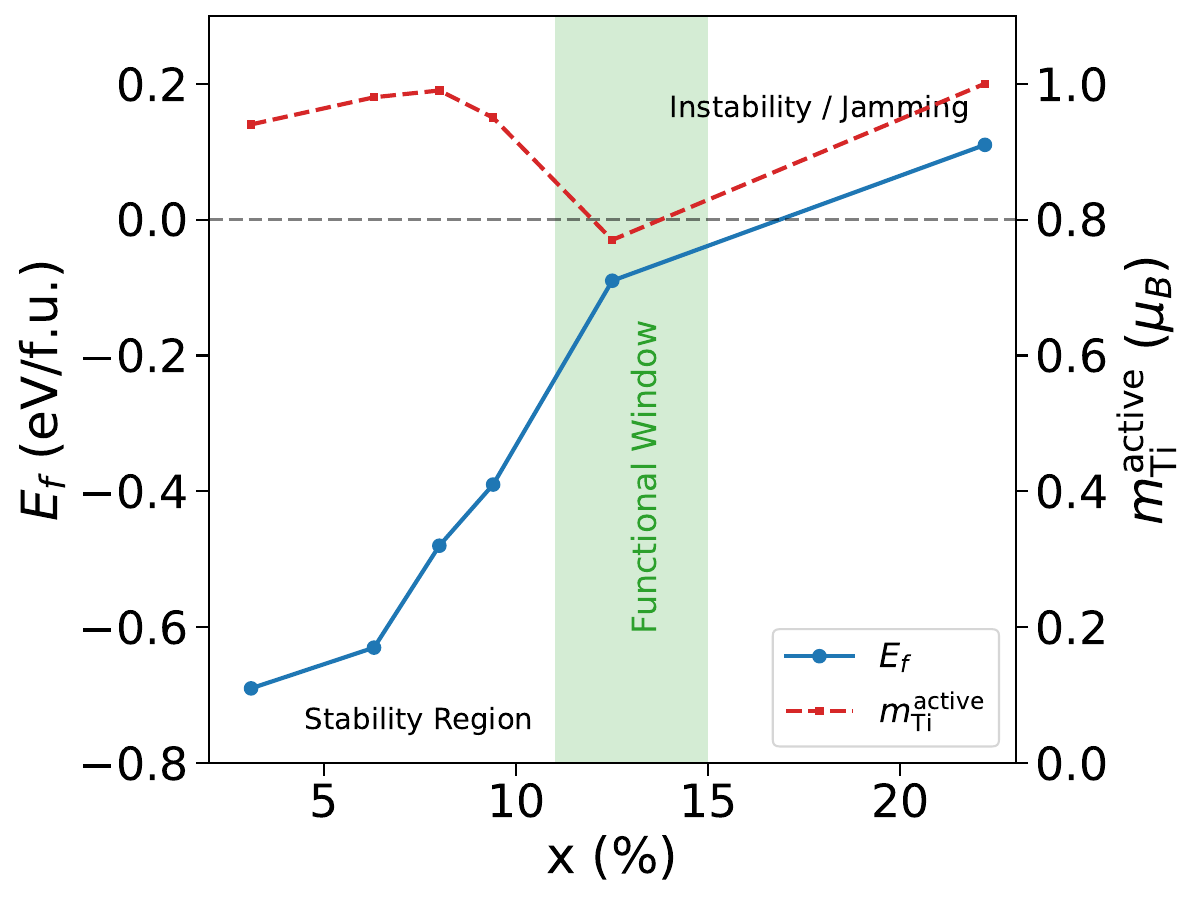}
		\caption{}
		\label{fig:thermo}
	\end{subfigure}
	\begin{subfigure}{0.32\textwidth}
		\includegraphics[width=\textwidth]{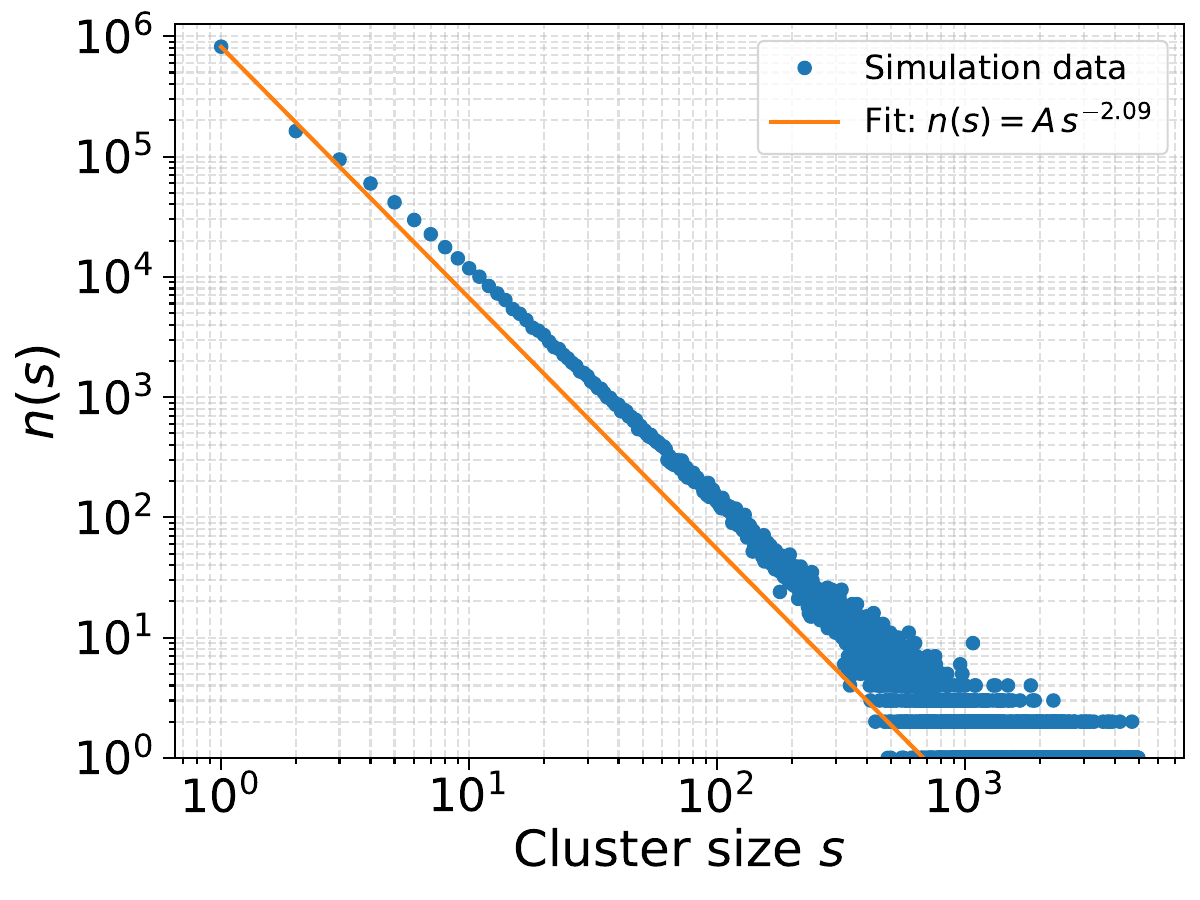}
		\caption{}
		\label{fig:tb}
	\end{subfigure}
	\begin{subfigure}{0.32\textwidth}
		\includegraphics[width=\textwidth]{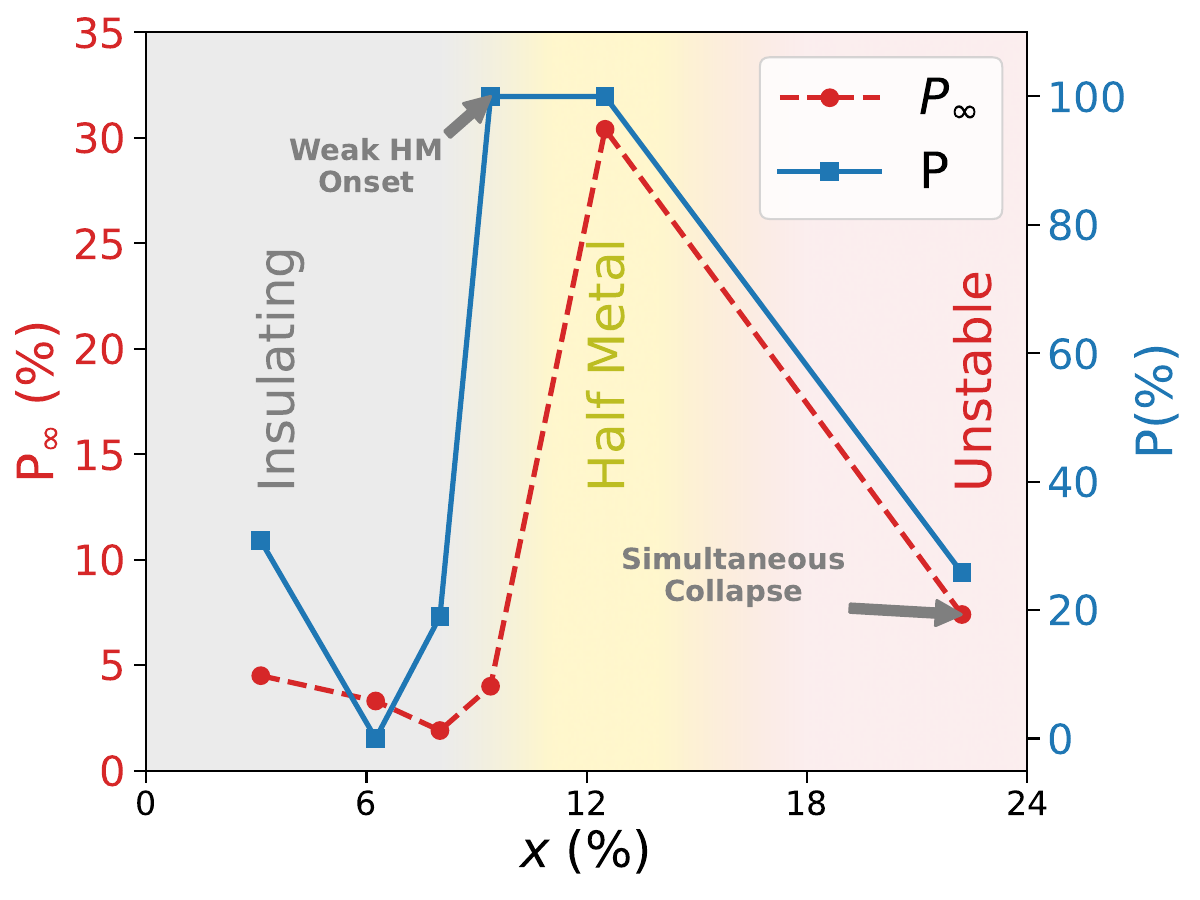}
		\caption{}
		\label{fig:phase}
	\end{subfigure}
	\caption{Stability and universality.
		(\subref{fig:thermo}) Thermodynamic limits: the defect formation energy ($E_f$) remains
		negative up to the jamming onset at $x \approx 20\%$.
		(\subref{fig:tb}) Universality: the cluster size distribution $n(s)$ from finite-size scaling
		($N = 6400$ sites) yields $\tau_{\text{TB}} = 2.09 \pm 0.03$, confirming the 2D percolation
		universality class.
		(\subref{fig:phase}) Phase diagram: the giant cluster fraction $P_\infty$ (red) undergoes a
		discontinuous jump at $x_c = 12.5\%$, perfectly synchronized with the onset of 100\% spin
		polarization (blue). Both order parameters collapse simultaneously at $x \approx 22\%$.}
	\label{fig:mechanism}
\end{figure*}

To establish the geometric origin of the half-metallic transition, we computed the giant cluster fraction ($P_\infty$)---the
probability that a vacancy belongs to the system-spanning cluster---across the full concentration range, and overlaid it with
the electronic spin polarization (\figref{fig:phase}).

The resulting phase diagram reveals a remarkable synchronization between topology and electronics. In the dilute regime
($x \le 9\%$), connectivity is negligible: $P_\infty < 5\%$ for all sub-critical configurations. A sharp geometric transition
occurs at $x = 12.5\%$, where $P_\infty$ undergoes a discontinuous jump to 30.4\%. This geometric percolation threshold coincides
precisely with the electronic onset of robust half-metallicity ($P = 100\%$) and the delocalization of the magnetic moment
(\tabref{tab:spinpol}), demonstrating that long-range ferromagnetic order is topologically gated by the spanning cluster.

To rule out finite-size artifacts, we fixed the vacancy count at $N_{\text{vac}} = 4$ and varied the supercell size. $P_\infty$
peaks at $4\times4$ (30.4\%), exceeding both the $5\times5$ (1.9\%) and $3\times3$ (7.4\%) configurations. This confirms that
the transition is concentration-controlled: the $5\times5$ cell at $x = 8\%$ is sub-critical, while the $3\times3$ cell at
$x = 22\%$ lies in the unstable, jammed regime. The thermodynamic stability boundary [Fig.~\ref{fig:thermo}] independently
confirms that the functional window closes at $x \approx 20\%$ where $E_f$ changes sign.

At higher densities ($x = 22\%$), both $P_\infty$ and the spin polarization collapse simultaneously. Despite a high local defect
density, the large Mean Cluster Size (16.7 sites) paired with a low $P_\infty$ (7.4\%) signals a transition from defect
percolation to phase separation: vacancies coalesce into dense, isolated droplets rather than extended networks.

\subsection{Critical Exponents and Universality Class}
\label{sub:scaling}

To classify the geometric transition rigorously, we employed two complementary approaches: large-scale tight-binding (TB)
simulations for precise exponent extraction, and direct analysis of the DFT charge-density clusters for \textit{ab initio}
verification.

The TB model ($N = 6400$ sites, 10 disorder realizations) yields a cluster size distribution $n(s)$ that follows a clean
power-law decay at the critical threshold [\figref{fig:tb}]. A least-squares fit gives the Fisher exponent
$\tau_{\text{TB}} = 2.09 \pm 0.03$ ($R^2 = 0.99$), in excellent agreement with the exact theoretical value for 2D percolation
($\tau_{\text{theory}} = 187/91 \approx 2.05$~\cite{stauffer1994introduction}). This quantitative match places the transition
firmly within a well-defined universality class, establishing a rigorous connection between the electronic phase transition in
this specific material and the broader theory of geometric critical phenomena.

\begin{figure}[htpb]
	\centering
	\begin{subfigure}{.9\columnwidth}
		\centering
		\begin{overpic}[width=.9\textwidth]{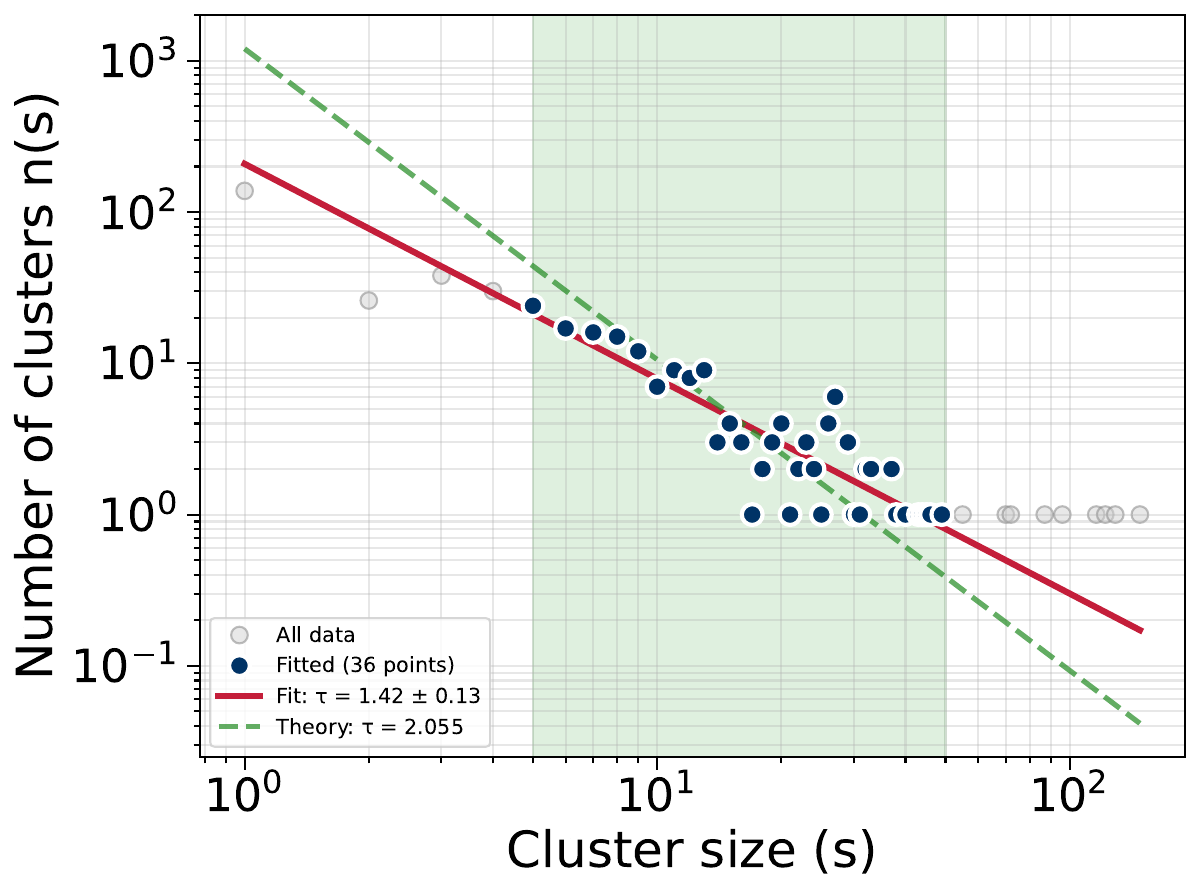}
			\put(18,15){%
				\colorbox{mplc7!30}{%
					\includegraphics[trim=4cm 4cm 6cm 4cm, clip,
						width=0.3\textwidth]{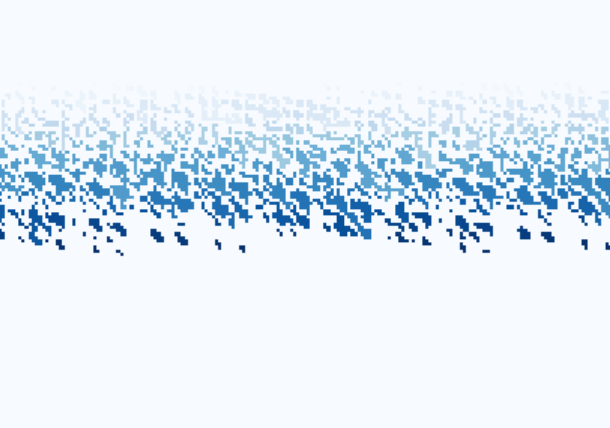}
				}
			}
		\end{overpic}
		\caption{}
		\label{fig:dftscale6}
	\end{subfigure}
	\begin{subfigure}{.9\columnwidth}
		\centering
		\begin{overpic}[width=.9\textwidth]{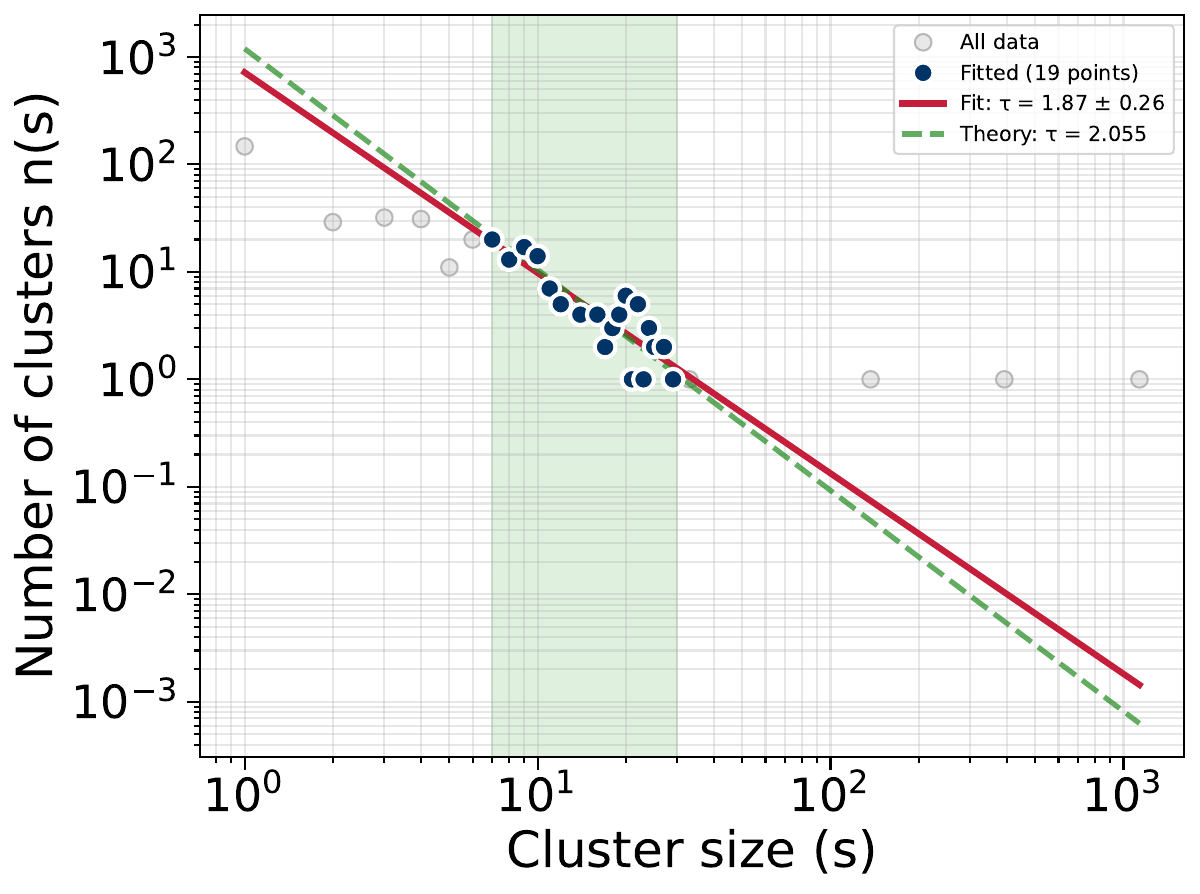}
			\put(60,48){%
				\colorbox{mplc7!30}{%
					\includegraphics[trim=4cm 4cm 6cm 4cm, clip,
						width=0.3\textwidth]{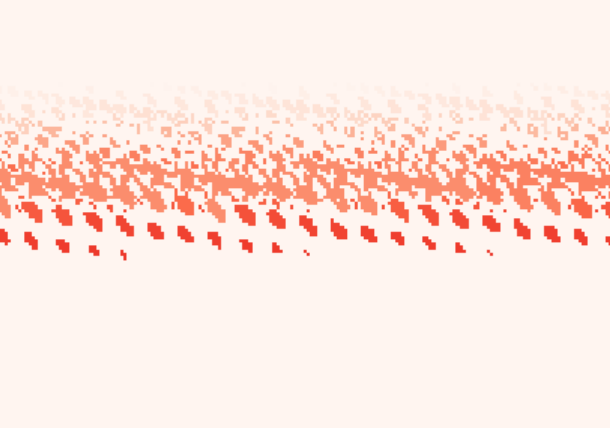}
				}
			}
		\end{overpic}
		\caption{}
		\label{fig:dftscale12}
	\end{subfigure}
	\caption{Evolution of geometric criticality in \textit{ab initio} charge densities.
		(\subref{fig:dftscale6}) Sub-critical regime ($x \approx 6.2\%$): the cluster size distribution
		(circles) yields a shallow effective exponent $\tau_{\text{eff}}^{\text{DFT}} \approx 1.42$.
		The deviation from the universal slope (dashed line) and the visual disconnectedness of the
		clusters (inset) indicate a finite correlation length.
		(\subref{fig:dftscale12}) Critical regime ($x \approx 12.5\%$): the magnetic polarons coalesce
		into a spanning network (inset). The distribution rectifies into a robust power law with
		$\tau_{\text{eff}}^{\text{DFT}} = 1.87 \pm 0.26$, converging toward the 2D percolation
		universality class ($\tau_{\text{theory}} \approx 2.05$, dashed line).}
	\label{fig:scaling}
\end{figure}

The fractal geometry encoded in the Fisher exponent is imprinted on the \textit{ab initio} electronic structure. We tracked
the renormalization of the effective exponent from the DFT charge-density clusters across the concentration range
(\figref{fig:scaling}). In the sub-critical regime ($x \approx 6.2\%$), the cluster size distribution deviates markedly from
power-law behavior, yielding a suppressed effective exponent $\tau_{\text{eff}}^{\text{DFT}} \approx 1.42$
[\figref{fig:dftscale6}]. The visual disconnectedness of the vacancy clusters (inset) confirms that the correlation length
$\xi$ remains finite and smaller than the system size, as expected below the percolation threshold.

At the half-metallic threshold ($x \approx 12.5\%$), the charge-density clusters coalesce into a spanning network (inset,
\figref{fig:dftscale12}), and the distribution rectifies into a robust power law with
$\tau_{\text{eff}}^{\text{DFT}} = 1.87 \pm 0.26$. The convergence of this exponent toward the universal value ($\tau \approx
	2.05$) signals the divergence of the connectivity length, confirming that the onset of half-metallicity is a continuous phase
transition driven by the percolation of magnetic polarons. The residual discrepancy between $\tau_{\text{eff}}^{\text{DFT}}$ and
$\tau_{\text{TB}}$ is consistent with finite-size corrections in the small DFT supercells ($\sim 30$ sites) relative to the
TB lattice ($6400$ sites).

\subsection{Geometric Delocalization in the Tight-Binding Model}
\label{sub:tb_results}

\begin{figure*}[htpb]
	\centering
	\includegraphics[width=0.8\textwidth]{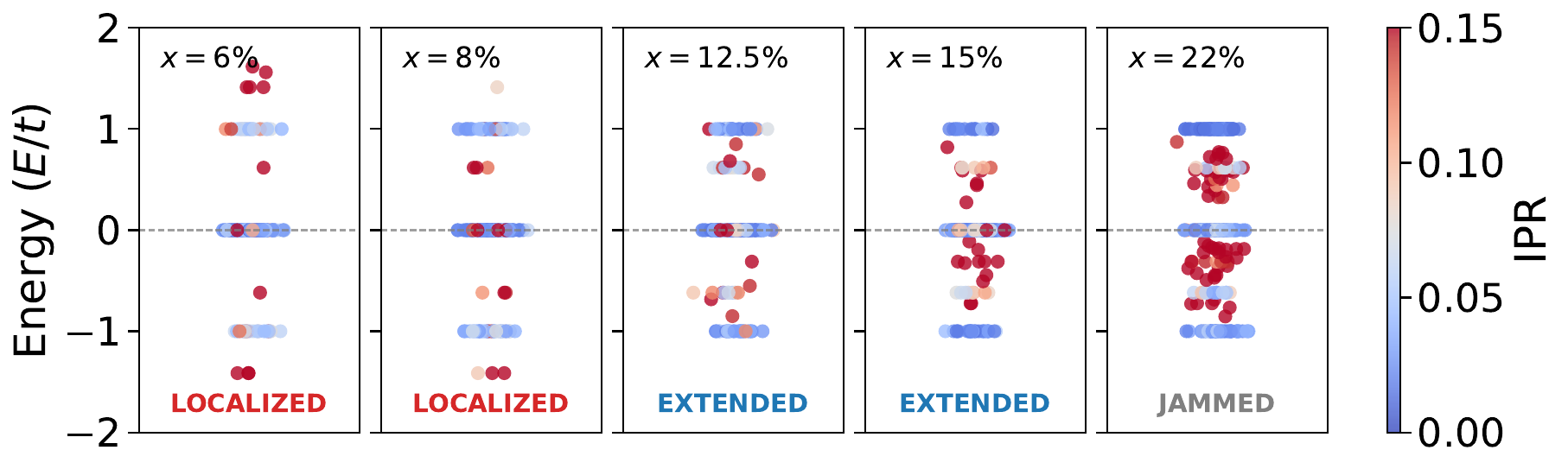}
	\caption{Geometric delocalization of the impurity band from tight-binding simulations
		($N = 3600$ sites). Each point is a single eigenstate; colour encodes the IPR, where red
		indicates spatial localization and blue indicates extended states.
		(a--b) Below threshold: all states are localized on isolated clusters.
		(c) At $x \approx 12.5\%$: a sudden IPR collapse signals extended-state formation.
		(d--e) Above threshold: the bandwidth broadens into a robust metallic band.}
	\label{fig:tb_localization}
\end{figure*}

The TB model independently confirms the percolation-driven mechanism through the Inverse Participation Ratio (IPR), which
distinguishes localized from extended electronic states (\figref{fig:tb_localization}).

In the sub-critical regime ($x < 10\%$), the spectrum is dominated by states with high IPR (red), confined to isolated vacancy
clusters. Although these states generate a DOS peak near $E = 0$ in the non-interacting model, their extreme localization implies
that on-site Coulomb repulsion ($U$) would split them into upper and lower Hubbard bands in a fully correlated picture, opening a
Mott gap consistent with the insulating DFT DOS. This resolves an apparent discrepancy between the TB and DFT descriptions: the
high spectral weight at $E = 0$ in the non-interacting TB model does not correspond to metallic transport.

At the critical concentration ($x \approx 12.5\%$), the IPR undergoes a sudden collapse, signalling the formation of extended
Bloch-like states that span the lattice (blue). Above threshold, the bandwidth broadens into a robust metallic band. This
delocalization transition occurs precisely at the percolation threshold, providing a model-independent confirmation that the
insulator-to-half-metal transition is controlled by the geometric connectivity of the vacancy wavefunction network, not by the
smooth accumulation of spectral weight.

\subsection{Thermal Stability of the Half-Metallic Phase}
\label{sub:thermal}

A prediction of ground-state half-metallicity is of limited practical value unless the magnetic order survives at finite
temperature. To assess this, we extracted the effective exchange coupling from the energy difference between FM and AFM
configurations in the $2\times2$ supercell, where the vacancy pair is spatially constrained and the two magnetic states are both
well defined. The substantial energy splitting ($\Delta E \approx 0.65$~eV per vacancy) reveals a strong nearest-neighbour
exchange interaction.

Mapping this energy scale onto a Heisenberg model within the Mean Field Approximation (MFA) for dilute systems:
\begin{equation}
	k_B T_C \approx \frac{2}{3}\, x\, \Delta E_{\text{exc}}
	\label{eq:tc}
\end{equation}
yields an estimated Curie temperature $T_C \sim 600$~K. MFA is known to overestimate transition temperatures, particularly in
low-dimensional systems where critical fluctuations and disorder effects are pronounced. However, even a conservative 50\%
reduction preserves $T_C > 300$~K, placing the magnetic ordering well above room temperature.

Two physical arguments underpin this estimate. First, the percolation threshold ($x_c \approx 12.5\%$) is a purely geometric
quantity set by the connectivity of the defect network, and is independent of temperature. Temperature affects the \emph{magnetic
	ordering} within the connected cluster, not the existence of the cluster itself. Second, the double-exchange mechanism that
stabilizes the FM state involves kinetic energy gain from carrier delocalization across the entire spanning cluster---a
collective effect that is intrinsically more robust against thermal fluctuations than localized superexchange, because destroying
itinerant order requires simultaneously localizing carriers throughout the macroscopic network.

We note that long-range ferromagnetic order in a strictly 2D isotropic Heisenberg system is forbidden at finite temperature by
the Mermin-Wagner theorem~\cite{Gibertini2019, Huang2017}. However, this restriction is lifted by magnetic anisotropy, which in
\ptis~arises from the combined effects of the vacancy-induced $C_{4v}$ crystal field and spin-orbit coupling on the Ti $3d$
states. The experimental observation of 2D Ising ferromagnetism in monolayer \ce{CrI3} with $T_C = 45$~K~\cite{Huang2017}
confirms that anisotropy-stabilized magnetic order is physically realizable in the 2D limit. The substantially larger exchange
coupling in \tis~($\Delta E \approx 0.65$~eV, compared to $\sim 2$~meV in \ce{CrI3}) suggests that the critical temperature in
our system could be considerably higher, although a rigorous determination would require explicit evaluation of the magnetic
anisotropy energy and Monte Carlo simulation on the percolating cluster geometry---a natural direction for future work.

\section{Discussion}
\label{sec:discussion}

\subsection{Quantitative Predictions of the Percolation Framework}
\label{sub:predictions}

While it may be qualitatively expected that sufficient vacancy doping will eventually produce metallic behaviour, the percolation
framework yields several quantitative predictions that go well beyond this intuition and constitute the central contribution of
this work.

First, the identification of a \emph{narrow} functional window ($11\% < x < 15\%$) provides a concrete design constraint that is
not accessible from electronic structure calculations alone. The lower bound is set by the percolation threshold
($x_c \approx 12.5\%$), below which carriers remain Anderson-localized on isolated clusters despite the presence of robust local
moments. The upper bound is set by a geometric jamming instability: at $x > 20\%$, vacancies coalesce into dense, compact
droplets rather than extended networks, fragmenting the conducting path and collapsing the half-metallic phase. This re-entrant
behaviour---where increasing the defect density \emph{destroys} rather than strengthens half-metallicity---is a non-trivial
consequence of geometric frustration.

Second, the supercell-size dependence of the magnetic ground state (Sec.~\ref{sub:magnetic}) demonstrates that the FM state is
stabilized by network topology, not merely by vacancy concentration. At identical doping ($x = 12.5\%$), the $2\times2$
supercell yields an AFM ground state while the $4\times4$ supercell yields FM order---a reversal that is incompatible with any
mechanism that depends only on the number of electrons per unit cell.

Third, the Fisher exponent ($\tau = 2.09 \pm 0.03$) places the transition in a well-defined universality class, establishing a
rigorous link between the electronic phase transition in this specific material and the broader theory of geometric critical
phenomena~\cite{stauffer1994introduction, saberi2015recent}. This universality implies that the mechanism is transferable to
other defect-doped 2D systems with similar extended-range coupling.

\subsection{Comparison with Other TMDC Systems}
\label{sub:comparison}

The percolation-driven mechanism identified here appears to be a general feature of octahedral TMDCs with spatially extended
defect states. Recent studies of \ce{VSe2}~\cite{Wang2023} and 1T'-\ce{MoS2}~\cite{Cai2015} report metallic onsets at
$x \approx 10$--$15\%$, consistent with our threshold. In contrast, $2H$-\ce{MoS2} remains semiconducting at comparable doping
levels~\cite{Gaikwad2025} because inward atomic relaxation around the vacancy quenches the magnetic moment before a percolating
network can form.

This contrast establishes a \emph{geometric selection rule}: only octahedral (1T-like) TMDCs whose defect wavefunctions extend
over $\sim 5$--$8$~\AA\ can support percolation-driven half-metallicity at experimentally accessible concentrations. In the
$2H$ polymorphs, the more localized defect states would require vacancy densities exceeding the thermodynamic stability limit to
achieve percolation. This selection rule provides a predictive criterion for identifying candidate materials among the broader
TMDC family, including isostructural compounds such as \ce{ZrS2} and \ce{HfS2} that share the $1T$ octahedral coordination and
similar Ti-group $d$-orbital characteristics.

Our findings also connect to the established literature on percolation in three-dimensional diluted magnetic semiconductors.
Bergqvist \etal~\cite{Bergqvist2004} demonstrated that ferromagnetism in (Ga,Mn)As requires Mn concentrations above the
percolation threshold for carrier-mediated exchange, and Sato \etal~\cite{Sato2010} showed that the magnetic phase diagram of an
entire class of DMSs is fundamentally controlled by percolation geometry. Stiller and Esquinazi~\cite{Stiller2023} extended this
reasoning to defect-induced ferromagnetism in \ce{TiO2}, identifying a quasi-2D percolation threshold. Our work carries this
framework into monolayer vacancy-doped systems, where the reduced dimensionality produces qualitatively new phenomenology: the
functional window is bounded from both sides---below by sub-critical localization and above by geometric jamming---a dual
constraint absent in the 3D DMS literature.

\subsection{Experimental Verification and Feasibility}
\label{sub:experimental}

The predictions of this work are experimentally testable with existing techniques and materials.

\textit{Materials platform.}---The clean limit of monolayer $1T$-\ptis~has been experimentally established: Yanagizawa
\etal~\cite{Yanagizawa_2025} synthesized the monolayer and characterized its band structure via angle-resolved photoemission
spectroscopy (ARPES), confirming the band gap and electronic structure that our DFT calculations reproduce. Controlled
vacancy engineering in TMDCs is a mature experimental technique. Electron irradiation~\cite{Komsa2012}, ion bombardment, thermal
annealing in vacuum, and chemical treatments~\cite{Hossen2024} have all been demonstrated to produce chalcogen vacancies at
controlled concentrations in \ce{MoS2}, \ce{WS2}, and related materials. The same methodologies are directly applicable to
\ptis.

\textit{Transport fingerprint.}---We predict that the anomalous Hall conductivity should exhibit a non-monotonic concentration
dependence: rising sharply at $x \approx 12.5\%$, peaking within the functional window, and collapsing beyond $x > 20\%$ due to
geometric fragmentation. This non-monotonic profile is a distinctive signature of the percolation mechanism that distinguishes
it from a simple monotonic increase expected from conventional band-filling arguments.

\textit{Spectroscopic fingerprint.}---The predicted 15-fold asymmetric broadening of the majority-spin impurity band (from
$W < 0.1$~eV to $W \approx 1.5$~eV) at the percolation threshold should be directly observable via spin-resolved ARPES. Scanning
tunnelling spectroscopy (STS) should reveal majority-spin gap closure ($\Delta E \lesssim 50$~meV) with sub-nanometre spatial
heterogeneity reflecting the fractal geometry of the spanning cluster. The minority-spin gap
($\Delta_\downarrow = 1.0$~eV) can be independently verified.

\textit{Spin polarization.}---The predicted 100\% spin polarization at $E_F$ can be measured using point-contact Andreev
reflection (PCAR), spin-polarized STM, or spin-resolved photoemission. A concentration-dependent measurement crossing the
percolation threshold would provide definitive evidence for the geometric mechanism.

\subsection{Implications for Spintronic Device Design}
\label{sub:devices}

The identification of a quantitative functional window ($11\% < x < 15\%$) translates directly into engineering design
parameters. The relevant materials figures of merit for device-level modelling are: 100\% spin polarization at $E_F$; a
minority-spin gap of $\Delta_\downarrow = 1.0$~eV, which determines the spin-filter efficiency and the thermal robustness of the
polarization; and an estimated Curie temperature exceeding 300~K, confirming the viability of room-temperature operation. These
parameters, together with the thermodynamic stability limit ($x < 20\%$), define the processing window within which reliable
spin-filter or spin-valve architectures could be constructed.

The geometric nature of the transition also opens a route to \emph{reconfigurable} spintronics. Because the half-metallic state
is controlled by the connectivity of the vacancy network rather than by bulk carrier density, local manipulation of vacancy
positions---for example, via focused electron beams~\cite{Komsa2012} or tip-induced vacancy migration in STM---could enable
the writing of spin-polarized conducting channels with nanometre-scale precision. This represents a conceptually distinct
paradigm from conventional gate-tunable spintronics, in which the spin polarization is modulated globally rather than patterned
spatially.

Non-equilibrium Green's function (NEGF) transport simulations on the percolating cluster geometry, incorporating the DFT-derived
band parameters reported here, would provide the quantitative current-voltage characteristics needed for device optimization. We
identify this as a high-priority direction for future work.

\section{Conclusion}
\label{sec:conclusion}

We have demonstrated that vacancy-doped monolayer \ptis~exhibits robust half-metallic ferromagnetism, resolving the long-standing
paradox of why defect engineering in 2D materials frequently produces local moments without itinerant magnetism. The resolution
lies in a two-step mechanism with distinct chemical and geometric prerequisites. Crystal-field symmetry breaking
($O_h \to C_{4v}$) upon sulfur removal selectively stabilizes the Ti $d_{z^2}$ orbital, generating robust paramagnetic
Ti$^{3+}$ ($d^1$) centers that are immune to the moment-quenching relaxation that plagues Group-VI TMDCs. However, these local
moments remain individually impotent: itinerant spin-polarized transport emerges only when the vacancy network percolates into a
system-spanning cluster that overcomes Anderson localization and activates double-exchange coupling.

Three quantitative results underpin this conclusion. First, half-metallicity with 100\% spin polarization and a minority-spin gap
of $\Delta_\downarrow = 1.0$~eV onset precisely at the percolation threshold $x_c \approx 12.5\%$, synchronized with the
formation of the giant cluster ($P_\infty \approx 30\%$). Second, finite-size scaling on $80 \times 80$ tight-binding lattices
yields a Fisher exponent $\tau = 2.09 \pm 0.03$, in excellent agreement with the exact 2D percolation universality class
($\tau_{\text{theory}} \approx 2.05$)---a result independently corroborated by the fractal scaling of \textit{ab initio} charge
densities ($\tau_{\text{eff}}^{\text{DFT}} = 1.87 \pm 0.26$). Third, a geometric jamming instability at $x > 20\%$ fragments
the percolating network and simultaneously collapses both thermodynamic stability and spin polarization. Together, these findings
define a narrow functional window ($11\% < x < 15\%$) bounded from below by sub-critical localization and from above by
geometric frustration---a dual constraint that demands precise experimental control over vacancy concentration and explains the
historical difficulty of achieving half-metallicity in doped TMDCs.

The emergence of universal critical exponents establishes that this percolation-driven mechanism is not specific to \ptis~but
belongs to a broader class of geometric phase transitions in defect-engineered materials. By connecting to the established
framework of magnetic percolation in three-dimensional diluted magnetic
semiconductors~\cite{Dietl2000, Bergqvist2004, Sato2010}, our work extends this paradigm to the 2D vacancy-doped regime, where
reduced dimensionality introduces qualitatively new physics---notably the upper jamming bound, which has no counterpart in the
3D DMS literature. The geometric selection rule that emerges---only octahedral TMDCs with spatially extended defect states
($\sim 5$--$8$~\AA) can support percolation-driven half-metallicity at accessible concentrations---provides a predictive
criterion for identifying candidate materials across the isostructural family, including \ce{ZrS2} and \ce{HfS2}.

With monolayer $1T$-\ptis~now experimentally accessible via ARPES~\cite{Yanagizawa_2025} and controlled vacancy engineering
well established in TMDCs~\cite{Komsa2012, Hossen2024}, the path to experimental verification is open. We have identified three
specific fingerprints: a non-monotonic anomalous Hall conductivity that peaks at the percolation threshold; a 15-fold asymmetric
broadening of the majority-spin impurity band observable via spin-resolved ARPES; and fractal spatial heterogeneity in
STM/STS mapping at the critical concentration. These predictions, together with the quantitative materials parameters reported
here (spin polarization, minority-spin gap, exchange coupling, and Curie temperature estimate), provide the inputs needed for
device-level transport simulations and targeted experimental synthesis. By shifting the design paradigm for 2D spintronics from
empirical chemical doping to quantitative geometric connectivity, this work opens a route to the rational engineering of
half-metallic states in defect-engineered van der Waals materials.

\section{Acknowledgments}
\label{sec:acknowledgement}
S.D.\ acknowledges the SRM Institute of Science and Technology (SRMIST) for a research fellowship. R.B.\ acknowledges financial
support from SRMIST under the Selective Excellence Research Initiative (SERI) grant. The authors thank the High-Performance
Computing (HPC) facility at SRMIST for providing computational resources. We also acknowledge the National Supercomputing Mission
(NSM) for facilitating access to the PARAM Smriti supercomputing facility.

\section*{Data Availability}
The data supporting the findings of this study are available from the corresponding author upon reasonable request.


%
\end{document}